\documentclass[technologies,perspective,accept,moreauthors,pdftex,10pt,a4paper]{Definitions/mdpi}

\firstpage{1}
\articlenumber{1}
\pubvolume{xx}
\pubyear{2018}
\copyrightyear{2018}

\history{Received: 28 September 2018; Accepted: 3 December 2018; Published: date}
\issuenum{1}

\usepackage{gensymb}

 \usepackage[labelformat=simple]{subcaption}

\DeclareCaptionLabelFormat{subcaptionlabel}{\normalfont(\textbf{#2}\normalfont)}
\usepackage{natbib}
\usetikzlibrary{shapes.arrows, shapes, shadows}

\newcommand{\bm}[1]{\boldsymbol{#1}}                 
\newcommand{\bset}[1]{\big\lbrace {#1} \big\rbrace}  
\newcommand{\ud}{\mathrm{d}}

\newcommand{\tder}[2]{\frac{\ud{#1}}{\ud{#2}}}
\newcommand{\integ}[2]{\int {#1}\ud {#2}}
\newcommand{\braket}[1]{\langle{#1}\rangle}

\newcommand{\Ron}{R_{\text{on}}}                     
\newcommand{\Roff}{R_{\text{off}}}                   

\DeclareMathOperator{\trace}{trace}

\graphicspath{{figures/}}

\Title{Memristors for the Curious Outsiders}

\Author{Francesco Caravelli  
$^{1,}$*\orcidA{} and Juan Pablo Carbajal $^{2,3}$\orcidB{}}

\AuthorNames{Francesco Caravelli, Juan Pablo Carbajal}

\address{%
$^{1}$ \quad Theoretical Division and Center for Nonlinear Studies, Los Alamos National Laboratory, \mbox{Los Alamos, NM 87545, USA}\\
$^{2}$ \quad  Institute  
for Energy Technology, HSR (Hochschule für Technik Rapperswil) or University of Applied Sciences Rapperswil, \mbox{8640 Rapperswil, Switzerland;} juan.pablo.carbajal@hsr.ch\\
$^{3}$ \quad Swiss Federal Institute of Aquatic Science and Technology, Eawag, Überlandstrasse 133, \mbox{8600 Dübendorf, Switzerland}}

\corres{Correspondence: caravelli@lanl.gov} 

\abstract{We present both an overview and a perspective of recent experimental advances and proposed new approaches to performing computation using memristors.
A memristor is a 2-terminal passive component with a dynamic resistance depending on an internal parameter.
We provide an brief historical introduction, as well as an overview over the physical mechanism that lead to memristive behavior.
This review is meant to guide nonpractitioners in the field of memristive circuits and their connection to machine learning and neural computation.}


\begin{document}

\section{Introduction}

The present review aims at providing a structured view over the many areas in which memristor technology is becoming popular.
As in many other hyped topics, there is a risk that most of the activity we see today will dissipate into smoke in the coming years.
Hence, we have carefully selected a set of topics for which we have experience and we believe will remain relevant when memristors move out of the spotlight.
After a general overview on memristors, we provide a historical overview of the topic.
Next, we present an intuitive and a mathematical view on the topic, which we trust is needed to understand why anybody would consider alternative forms of computation.
Thus, experts in the fields might find this article slow-paced.

\begin{figure}[H]
    \centering
    \includegraphics[width=0.3\textwidth]{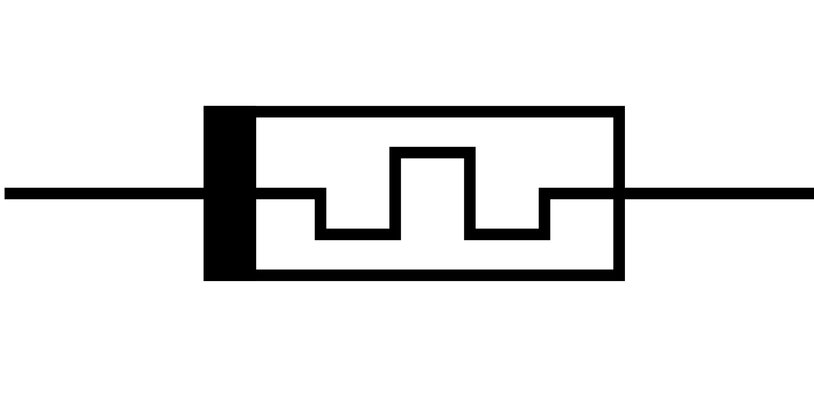}
    \caption{Electronic symbol of the memristor.}
    \label{fig:symbol}
\end{figure}

The memristor was introduced as a device that ``behaves somewhat like a nonlinear resistor
with memory''~\citep{Chua71}, the resistance of the component depends on the history of the applied inputs: voltage or current.
Different curves of the applied input elicit a different dynamic response and final resistance of the memristor.
In addition, if we remove the input after certain time, leave the component alone, and come back to use it, the device will resume its operation from a resistance very similar to the one in which we left it; that is, they act as non-volatile memories.
Furthermore, memristors also react to the direction of the current, i.e., they have polarity.
This polarity is shown explicitly on the symbol of the memristor used in electronic diagrams: a square signal line (as opposed to a zigzag line for the resistor), inside an asymmetrically filled rectangle, as in Figure~\ref{fig:symbol}.
The interplay between the response behavior and the non-volatility of the device defines its usability either as a storage device or for more involved purposes such as neuromorphic computing.
The present article gravitates around the fact that memristors are electronic components which behave similarly to human neuronal cells, and are an alternative building block for neuromorphic chips.
Memristors, unlike other proposed components with neural behavior, can perform computation without requiring complementary metal--oxide--semiconductor (CMOS) if not for readout reasons.

After the experimental realization of a memristor by~{Strukov et al.}, 
which brought them to their current popularity, Leon Chua wrote an article titled ``If it's pinched it's a memristor''~\citep{Chua2014}.
The~title refers to the fact that the voltage drop across the device shows a hysteresis loop pinched to the zero of the voltage--current axes when controlled with a sinusoidal voltage or current.
The claim by~Chua  is that a device which satisfies this property must necessarily be a memristor.
Chua also proved that such a device cannot be obtained from the combination of nonlinear capacitors, inductors and resistors~\citep{Chua71,Chua1976,Chua2014}.
It has been shown, however, that this property is a modeling deficiency for redox-based resistive switches~\citep{Valov2013}.

Resistive switching was of interest even before the 2008 article.
For instance, the review of~of Waser
and Aono  shows that Titanium Dioxide had been in the radar for non-volatile memories for decades.
Nevertheless, one can identify a clear explosion of interest after the 2008 publication.
Next, we give a list of established and new companies developing memristors technology using a variety of compounds.
The list is given without a particular order as these companies work on various types of compounds and type of memristors (for instance both resistive random-access memory (ReRam) and phase-change modulation (PCM)). 
Specifically, we are aware of: {HP, SK Hynix, IBM, IMEC, Fujitsu, Samsung}, SMIC, Sharp, TSMC, NEC, Panasonic, Macromix, Crossbar Inc., Qimonda, Ovonyx, Samsung, Intel, KnowM, 4DS Memories Ltd., Global Foundaries, Western Digital (before called SanDisk), Toshiba,  Macronix, Nanya, NEC, Rambus, ST Microelectronics, Winbond, Adesto Technologies Corporation, HRL Laboratories LLC, and Elpida. 

Going into the many physical mechanisms that make a memristive device work does require a deep knowledge of material properties.
For example, the hysteretic behavior can be caused by Joule heating, as shown in an example taken from macroscopic granular materials~\citep{Bequin2010}.
In addition, hysteresis is a phenomenon that is common in nanoscale devices and it can be derived using Kubo response theory~\citep{DiVentra2013}.
Kubo response theory is used to calculate the correction to the resistance induced by a time dependent perturbation: this is the formal framework to address resistive switching due to either electrical, thermal or mechanical stresses in the material.
We can classify physical mechanisms that lead to memristive behavior in electronic components in four main types:

\begin{itemize}[leftmargin=*,labelsep=5.8mm]
    \item Structural changes in the material (PCM like): in these materials, the current or the applied voltage triggers a phase transition between two different resistive states;
    \item Resistance changes due to thermal or electric excitation of electrons in the conduction bands (anionic): in these devices, the resistive switching is due to either thermally or electrically induced hopping of the charge carriers in the conducting band.
For instance, in Mott memristors, the resistive switching is due to the quantum phenomenon known as Mott insulating-conducting transition in metals, which changes the density of free electrons in the material.
    \item Electrochemical filament growth mechanism: in these materials, the applied voltage induces filament growth from the anode to the cathode of the device, thus reducing or increasing the~resistance;
    \item \textls[-15]{Spin-torque: the quantum phenomenon of resistance change induced via the giant magnetoresistance} switching due to a change in alignment of the spins at the interface between two differently polarized magnetic materials.
\end{itemize}

 These mechanisms above are truly different in nature, and whilst not the only ones considered in the literature, are the most common ones.
We provide a technical introduction to these mechanisms in Appendix~\ref{sec:appphys} for completeness.

The primary application of memristors, as we will show in this article, is towards neuromorphic computing.
The word neuromorphic was coined by Carver Mead~\citep{Mead1990} to describe analog circuits which can mimic the behavior of biological neurons.
In the past several years, the field has experienced an explosive development in terms of manufacturing neuromorphic novel chip architectures which can reproduce the behavior of certain parts of the brain circuitry.
Among the components used in neuromorphic circuits, we consider memristors whose behavior resembles the one of a certain type of neurons.
The~analogies between biological neuronal systems and electronic circuits are manifold: conservation of charge, thresholding behavior, and integration to mention just a few.
For instance, diffusion of calcium across the membrane can be mapped to diffusion in electronic components, and a computational role associated to the circuit design.
In addition, it is known that the brain is power efficient: roughly twenty percent of an individual's energy is spent on brain activity, and this is roughly around 10 W.
Besides their role in biological neural models (Section~\ref{sec:memgalore}), we also discuss theoretical approaches to memristive circuits and their connection to machine learning (Sections~\ref{sec:memstorage} and~\ref{sec:memdataproc}).

The connection to machine learning is complemented with an overview on analog computation (Section~\ref{sec:analogc}).
Historically, the very first (known) computer was analog.
It is (approximately) 2100 years old, and it has been found in a shipwreck off the cost of the island of Antikythera at the beginning of the past century~\citep{antmec} (only recently, it has been understood as a model of planet motion).
Despite our roots in analog computation, our era is dominated by digital computers.
Digital computers have been extremely useful at performing several important tasks in computation.
We foresee that future computers will likely be a combination of analog (or quantum analog) and digital (or quantum digital) computing chips.
At the classical level, several analog computing systems have been proposed.
Insofar, most of the proposed architectures are based on biological systems, whose integration with CMOS can be challenging, and this is the reason why analog electronic components are seen as promising alternatives~\citep{Adamatzky2016,Dalchau2018}.

Digital computation has been dominated by the von Neumann architecture in combination with CMOS technology.
Within this architecture, we find two types of memory: Random Access Memory (RAM), a volatile and quick memory in which computation is performed, and non-volatile Hard Disk (HD) for long-term storage of information.
The neat separation between memory (RAM and HD) and computing (the processors or central processing unit (CPU)) is a key feature of computation using von Neumann architecture. 
It requires that the data is split into packets, transferred to the CPU where computation is performed, and then repeated until the full computation task has been completed.
As far as our understanding goes, this separation is not present in the brain, in which memory and processing happen within the same units.
Several proposed architectures that mimic this property have been proposed, most of them based on memristors~\citep{DiVentra2013b}.
This type of computation is referred to as memcomputing.

This article is organized as follows:
We first provide a historical introduction to memristors, as it is understood by the authors (Section~\ref{sec:history}).
We then provide the key ideas behind the technology with simple mathematical models (Section~\ref{sec:models}).
Albeit separated from the main text, we have provided an introduction to the main technology and physical principles underlying memristors in Appendix~\ref{sec:appphys}.
We then focus our gaze on the description and use of memristors both for data storage (Section~\ref{sec:memstorage}) and data processing (Section~\ref{sec:memdataproc}): the former is the current target for marketing the technology, the second is believed to be the main application of memristors in the long run.
The similarity between memristors and neurons allows the implementation of machine learning on chip via Memristor/FPGA (field-programmable gate array) 
interfaces using crossbar arrays, we cover this topic in Section~\ref{sec:crossbar}.
We have dedicated Section~\ref{sec:analogc} to overview the fundamental topic of analog computation, followed by a brief recapitulation of machine learning techniques that can be seamlessly used with memristors (Sections~\ref{sec:reservoir}--\ref{sec:NEF}).
We discuss in Section~\ref{sec:memgalore} computation as an epiphenomenon of memristor dynamics and in connection with CMOS.
We close the article with some remarks that should help in the discussion of the topics covered.

\section{Brief History of Memristors}
\label{sec:history}
The earliest known occurrence of a ``memristive'' behavior in circuit components is in the study, by Sir Humprey Davy's, of arc (carbon) lamps~\citep{Davy}, dated to the late 19th century.
Another example, which was key to the discovery of the radio, is the coherer~\citep{Falcon2004,Falcon2005}.
The coherer was invented by Branly~\citep{Branly} after the studies of Calzecchi-Onesti, and inspired Marconi's radio receiver~\citep{Marconi}.
Branly's coherer serves as perfect homemade memristor~\citep{Bequin2010}, as it simply requires either a (fine) metallic filling or some metallic beads, and it falls within the Physics discipline of electrical properties of granular media.
Coherers are very sensitive to magnetic fields, and thus can act as a radio receiver and as we describe in Appendix~\ref{sec:appphys}. They do posses a typical hysteric behavior which is associated with memristive components.
At the simplest level of abstraction, a memristor is a very peculiar type of nonlinear resistance with an hysteresis (e.g., they possess an internal memory).
Currently, the discussion is focused not only on the application of this technology to novel computation and memory storage, but also on the fundamental role of memristors in circuit theory~\citep{Abraham2018a}.
While this discussion is important, we first discuss why memristive behavior is not uncommon in analog computation.
We thus use the analog of hydraulic computers.

Hydraulic computers were built in Russia during the 1930s before valves and semiconductor were invented.
The hydraulic computers designed by the Russian scientist Vladimir Lukyanov were used to solve differential equations; 
other models like the Monetary National Income Analogue Computer (or MONIAC)~\citep{Strogatz2009}, would be 
later employed by the Bank of England to perform economic forecasts.


\textls[-15]{The idea behind these computers was to use hydraulics to solve differential equations.
As Kirchhoff} laws can be stated for any conservative field, we have that the pressure drop in a loop is equal to the actions of pumps in that loop.
The conservation of mass is equivalent to conservation of charge, and to Kirchhoff's second law: in the steady state, the mass of water entering a node must be equal to the mass flowing out.
Ohm's law is equivalent to Poiseuille's law in the pipes: a porous material in the pipes, or a constriction, is equivalent to a resistance.
A material that expands when wet, like a sponge, will increase the resistance to the flow as it absorbs the water.
This is equivalent to a higher resistance which depends on the amount of water that flowed thought the system, i.e.,  memristor-like.
The sponge, however, does not have a polarity, while memristors do, depending on the physical mechanism which induces the switching.
Other memristor-like systems can be built with other mechanical analogs~\citep{Vongehr2015,Vongehr2015b}, plants and potato tubes~\citep{Volkov} or slime moulds~\citep{Gale2015,Gale2015a} just to name a few.

The modern history of memristors is tied to the work of Leon Chua. The first time the word memristor (as an abbreviation for memory resistor) appeared was in the now celebrated Chua 1971 article~\citep{Chua71}.
During the 1960s, Leon Chua worked extensively on the mathematical foundations of nonlinear circuits.
When he moved to Berkeley, where he currently is a Professor, he had already won several awards such as the IEEE ( Institute of Electrical and Electronics Engineers) Kirchhoff Award.   
The definition of memristor was made clearer in a second paper with his student at the time, Sung~Mo~Kang~\citep{Chua1976}.
The second work was an important generalization of the notion of memristor and is the one we used in the present paper.
Chua and Sung Mo Kang  introduced the notion of ``memristive device'': a resistance which depends on a state variable (or variables), which is sufficient to describe the physical state (resistance) of the device at any time.
The component defined by~Chua, and then by~Chua and Sung Mo Kang, is a resistance whose value depends on some internal parameter, which in turn has to evolve dynamically according either to current or voltage.
Implicitly, one needs to also define the relationship between the resistance and the state variable, which characterizes the device resistance.
In the analogy with hydraulic computers, the state variable represents by the density of holes in the sponge as a function of time, while the resistance is the amount of traversable area.
The 1976 and the 1971 papers were mostly mathematical and formal, without any connection to the physical properties of a real device.
The 1976 paper also introduced the fact that, if one controls the device with a sinusoidal voltage, then one should observe Lissajous figures in the current--voltage (I--V) diagram of the device.
It also established that any electronic component that displayed a pinched hysteresis in the I--V diagram has to be a memristor.
The eager reader will find more details on the devices in the rest of the paper.

For the 30 years after the work by~Chua and Sung Mo Kang, the field of memristors was basically non-existent.
During the mid 2000s, Strukov, Williams and collaborators, working at Hewlett-Packard Labs (Palo Alto, CA) 
were studying oxide materials and resistive switching. The physics of resistive switching has been known since the early  1970s, with the introduction of Phase-Change Materials.
The physical origin of resistive switching was well studied, albeit not fully understood~\citep{Waser2007,Waser2007b,Aono2010}.
The idea that memristors could be experimentally realized became popular with the article of~Strukov et al. at HP Labs.

\section{Mathematical Models of Memristors}
\label{sec:models}

In this section, we provide a general mathematical description of memristive systems followed by the details of the memristor model with linear memory dynamics proposed by~Strukov et al.  
This is one of the simplest memristors models that captures the behaviors relevant for technological applications of memristive systems.

The fundamental passive electric components are the resistors, capacitors and inductors.
These components couple the voltage difference $v$ applied between their ports with the current $i$ flowing through, with the following differential relationships:
\begin{gather}
  \begin{aligned}
    \text{Resistor:}  & \quad \ud v=R \, \ud i;\\
    \text{Capacitor:} & \quad \ud q = C \, \ud v, \quad \text{charge: } \ud q = i \, \ud t ;\\
    \text{Inductor:}  & \quad \ud\Phi=L \, \ud v, \quad \text{flux: } \ud \Phi = v \, \ud t.
  \end{aligned}
\end{gather}

\begin{figure}[H]
    \centering
    \includegraphics[width=0.7\textwidth]{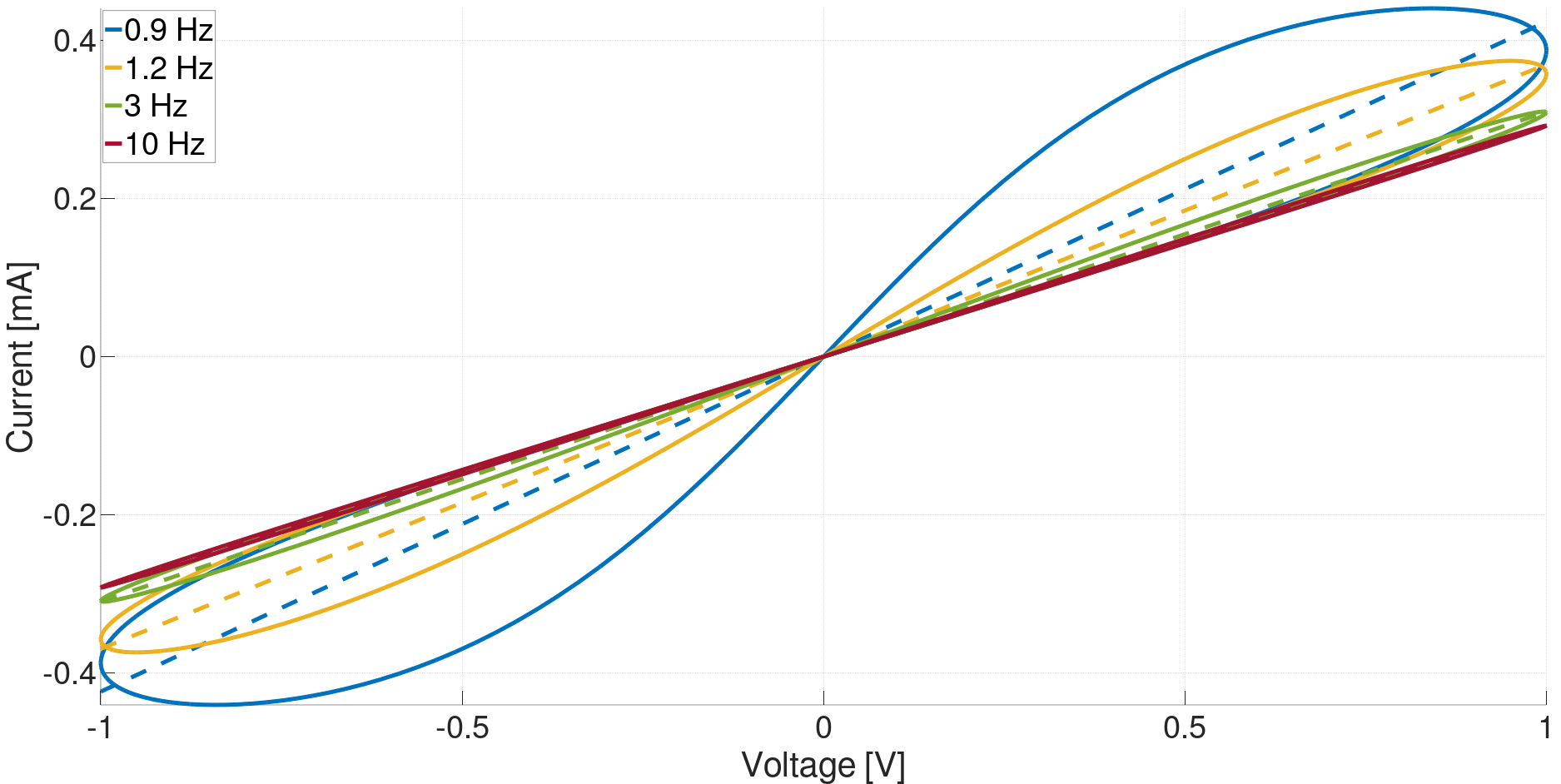}
    \caption{Pinched hysteresis loop of an ideal memristor.
    The parameters of the model are $\alpha = 0$, \mbox{$\beta$ = {0.3} {mA}}, $\Ron$ = {1}~{k$\ohm$} and $\Roff$ = {6}~k$\ohm$.     The driving voltage is $V(t) = \sin (2 \pi f t)$ with $f$ taking several values.
    We see that, for increasing frequencies, the hysteresis is reduced, eventually converging to a  line (resistor).
    The loop slope (dashed lines) decreases with increasing frequency.}
    \label{fig:pinchys}
\end{figure}

 These relations introduce the resistance $R$, the capacitance $C$, and the inductance $L$.
The ideal memristor was initially and abstractly formulated as the flux--charge relationship:

\begin{equation}
    \ud \Phi = M\, \ud q,
\end{equation}

\noindent which, in order to be invertible, must have $M$ defined either in terms of $\Phi$, or $q$, or it must be a constant.
The latter case coincides with the resistor, while if $M$ is defined in terms of $\Phi$, then we have a voltage-controlled memristor, if it is defined in terms of $q$ it is a current-controlled memristor. For the sake of simplicity we do not make a difference between current-controlled and charge-controlled; this is an abuse of the language. 
The units of the coupling $M$ are Ohms, as can be seen by the units of the quantities it relates, which justifies the notion of memristor at least as a nonlinear resistor.  
However, according to~Chua, only nonlinear resistors showing a pinched hysteresis loop, i.e., $V=0 \rightarrow I=0$, are classified as memristors.
An example is shown in Figure~\ref{fig:pinchys}.
Further classifications~(ideal, extended, generic, see \citep{Chua2014}) can be added to a device depending on other properties of the current--voltage ($I- V$) diagram that are observed experimentally when the device is controlled with a sinusoidal voltage at a certain frequency. 
For example, if for high frequencies the $I-V$ becomes a straight line (recovery of pure linear resistance, no hysteresis), the device is a generic memristor, under~Chua's nomenclature.
General memristors might be locally-active, showing regions where the resistance changes sign~\citep{Chua2015}.

Many systems, not necessarily electronic ones, can fulfill these properties.
Physical systems with memristor-like behavior are denominated memristive systems, and are described by the following non-autonomous (input-driven) dynamical system:
\begin{gather}
  \begin{aligned}
    \tder{\bm{x}}{t} &= F(\bm{x}, u),\\
    y &= H(\bm{x}, u) u,
  \end{aligned}
  \label{eq:eqmem}
\end{gather}

\noindent where $\bm{x}$ is a vector of internal states, $y$ a measured scalar quantity and $u$ a scalar magnitude controlling the system.
The pair $(y,u)$ is usually voltage--current or current--voltage, hence the second equation is simply Ohm's law for the voltage--current relationship.
In the first case (current-controlled system), the function $H(\bm{x},u)$ is the called memristance, and it is called memductance in the second case (voltage-controlled).
The function $H(\bm{x},u)$ is assumed to be continuous, bounded, and of constant sign. In order to rule out locally active devices, it is necessary to further require that $H(x,u)$ is monotonic in $u$. 
This leads to the zero-crossing property or pinched hysteresis: $u = 0 \Rightarrow y = 0$.
The~states $\bm{x}$ can be many physical states other than charge, e.g., the internal temperature of a granular material.
The~minimum number of internal states on which $H(\bm{x},u)$ depends is called the \textit{order} of the memristive~system.  

The reader should not overlook the difference between an ideal memristor (a mathematical model), memristive systems (also a mathematical model), and a memristor (a physical device).
Ideal memristors are models of devices in which the internal parameter controlling the resistance is the charge, which is linked to the flux.
Memristive systems generalize this behavior, and the internal parameters can be other physical quantities, or combination of physical quantities, even if they do not include flux or charged particles.
The latter is better attuned to many memristors, i.e., physical realizations of memristive behavior.

Among the (large) class of memristive systems, the model proposed by~Strukov et al. is appealing due to its simplicity.
This model, shown in Equations~\eqref{eq:memristivedyn} and \eqref{eq:memristivestate}, captures the basic properties of memristors that are relevant for the understanding of the device and for its technological applications.

Equations~\eqref{eq:memristivedyn}  and \eqref{eq:memristivestate} model the behavior of a current-controlled Titanium Dioxide memristor~\citep{Strukov2008}, also known as the HP-memristor:
\begin{align}
    \tder{w(t)}{t} &= \alpha w(t)-\frac{1}{\beta} I(t), \label{eq:memristivedyn}\\
    \frac{V(t)}{I(t)} &= R\left(w(t)\right) := \Ron \left[1 - w(t)\right] + \Roff w(t), \label{eq:memristivestate}\\
    w(0) &= w_0 \rightarrow R(w_0) = R_0, \quad  0 \leq w(t) \leq 1,
\end{align}

\noindent where $I(t)$ is the current flowing in the device at time $t$.
The quantities $\Ron \leq \Roff$ in the parametrization of the resistance $R$ in terms of the adimensional variable $w(t)$ (called \textit{memory}), define the two 
limiting resistances.
The parameters $\alpha$ (with units s$^{-1}$) and $\beta$ (with units s$^{-1}$A$^{-1}$), define the dynamic properties of the memristor.
The parameter $\alpha \geq 0$, sometimes called \textit{volatility}, indicates how fast the resistance drifts towards $\Roff$ in the absence of currents~\citep{volc}.  
Since the resistance function depends on a single state $w$, this is a first order memristive system.
Although this model has several drawbacks when it comes to its connection to the physical behavior of ion migration in the conductor~\citep{DiVentra2013,Abraham2014,Vongehr2015,Abraham2016,Abraham2018a,Abraham2018,Tang2016}, it will be the reference for most derivations and discussions in this article, as it captures several key properties of a generic memristor.

We begin with the case with $\alpha=0$, i.e., a non-volatile memristor.
We solve the system of equations above using a zero mean small amplitude driving voltage $V(t)$ which is such that, for all times $0< w(t)< 1$, i.e., the memristor should not saturate at any time~\citep{Wang}:

\begin{equation}
    R\left(w(t)\right)\tder{w(t)}{t}=\tder{}{t} \left( (\Roff-\Ron) \frac{1}{2} w^2(t)+ \Ron w(t) \right)=- V(t),
\end{equation}

\noindent from which we derive, if we define $\xi=\frac{\Roff-\Ron}{\Ron}$ and integrate over time:

\begin{equation}
 \left( \frac{\xi}{2} w^2(t)+ w(t)\right)= \left( \frac{\xi}{2} w^2(t_0)+ w(t_0)\right)+ \int_{0}^t V(\tau) \ud \tau, \label{eq:periomemsol}
\end{equation}

\noindent whose solution is given by

\begin{equation}
    w(t)=\frac{\sqrt{2 \left(\frac{\xi}{2} w^2(t_0)+ w(t_0)\; + \; \int_{0}^t V(\tau) \ud \tau\right) \xi +1}-1}{\xi }.
\end{equation}

This equation fulfills the fundamental property of a memristor: it has the zero crossing property (pinched hysteresis loop).
Furthermore, it shows properties of generic memristors: the loop tilts to the right for driving signals with increasing frequencies, becoming a straight line for sufficiently high frequencies.
To see the latter, consider $V(t) = V_0 \cos \left(\omega t\right)$, as the frequency $\omega\rightarrow \infty$.
For this voltage, the integral in Equation~\eqref{eq:periomemsol} goes to $0$, and $w(t)\rightarrow w(t_0)$.
This implies a constant resistance, i.e.,~the~memristor becomes a resistor for high frequencies.

In this model for $\alpha=0$, we have that $w(t)\sim q(t)$, where $q(t)$ is the charge in the conductor.
For a titanium dioxide thin film, it was shown in~\citep{Strukov2008} that

\begin{equation}
    \beta^{-1} \sim \frac{\mu_e \Ron}{D},
\end{equation}

\noindent and thus

\begin{equation}
R(q) \approx \Roff \left( 1 - \frac{\mu_e \Ron}{D^2} q \right),
\end{equation}
 where $\mu_e$ is the electron mobility in the film, $\Roff$ is the resistance of the undoped material (for instance titanium oxides, $\sim${100}~${\ohm}$), and $D$ is a characteristic length of the film.
From the micrometer to the nanometer, the value of $\beta$ grows by a factor of $10^6$, which is why the memristance in this material is a nanoscale effect.

Memristors are also referred to as resistive switches, because when $\frac{\Roff}{ \Ron}\gg 1$ and $\beta$ is small enough, the memristor can be quickly driven from one state to the other via a current (or voltage) inversion.
The~operation is often referred to a $SET$ or $RESET$ 
 in the technical literature, depending on the operation one is interested in, and it allows the use of non-volatile memristors as bits.

In the volatile case, i.e., $\alpha > 0$, the memristor does not retain its memory state.
When a volatile memristor is not activated via an external forcing, the memristor drifts to its insulating state, \mbox{$R(t=\infty)=\Roff$.}
That is, if $I(t)=0$, then the internal memory is simply given by $w(t)= w_0 e^{\alpha t}$, where $w_0$ is the initial condition.
This behavior was first experimentally observed in~\citep{Ohno2011}.
Volatility is discussed again in Section~\ref{sec:volatility}.

Following the distinction between memristive and memristors we have discussed, when $\alpha \neq 0$, the parameter $w$ cannot be uniquely identified with the charge, hence the model does not correspond to a bona fide memristor.
In fact, for a single component the solution of:

\begin{equation}
    \tder{w(t)}{t}=\alpha w(t)-\frac{1}{\beta} \tder{q(t)}{t}
\end{equation}

\noindent is given by

\begin{equation}
    w(t)=\frac{e^{\alpha  t}}{\beta} \int_1^t e^{-\alpha  \tau } \frac{d}{d\tau}q(\tau ) \, d\tau +w_0 e^{\alpha  t},
\end{equation}

\noindent which is a composite function that includes the charge.

The numerical model of the memristor allows us to study their theoretical capabilities as well as simulating large networks of these devices.
From the point of view of numerical simulations, the~dynamics at the boundaries of $w$ need to be stable, or alternatively, $w$ be constrained in $[0,1]$.
In~general, the latter does not scale well in terms of runtime, and for this reason it is often useful to bound the internal states of the memristor model via the introduction of window functions~\citep{Pershin13ventra, Biolek2009, Biolek2013, Biolek2015, Nedaaee2011, Strukov2008, Joglekar2009}.
Since we are not aware of any systematic validations of windowing functions with real devices, we believe they should be considered tricks useful for large simulations.
It is common to use windowing techniques based on polynomials; for an extensive review, we refer the reader to~\citep{Biolek2015}.
Polynomial windowing functions that reproduce nonlinear dopants drift can be introduced via the so-called Joglekar window function $F(x)$~\citep{Joglekar2009, Biolek2009}, such that $F(1)=F(0)=0$, which generalizes the evolution of $w(t)$ as

\begin{equation}
    \tder{w(t)}{t} = \alpha w(t) - F(w) \frac{1}{\beta} I(t),
\end{equation}
\noindent where the window function can take the form $F(w) =1 - (2 w - 1)^{2p}$, with $p$ a positive integer.
Depending on the type of memristor model, different window functions might provide a better approximation of the nonlinear effects near the boundaries of the memory.
For an overview of the various physical mechanisms that can be involved, we refer the reader to Appendix~\ref{sec:appphys}.

Numerical simulations of memristive networks can be achieved with different methods.
Efficient ones include the SPICE methodology~\citep{spicememr,spicememr2} and the more recent Flux-Charge method~\citep{fluxcharge}.
Furthermore, general solvers for Differential-algebraic system of equations (e.g., SUNDIALS solvers~\citep{hindmarsh2005sundials}) can also be used for networks with a few hundred components, without major loss of performance.
For some networks (without capacitance or inductance), it is possible to derive a monolithic system of ordinary differential equations for the evolution of the network states, which already include the nonlocal algebraic constraints imposed by Kirchhoff's laws~\citep{Caravelli2017,Caravelli2017b}.
The choice of the method is largely based on the questions to be addressed by the simulations, which often results in a trade-off between generality of the solver (e.g., flexible memristor model) and performance.

\section{Memristors for Storage}
\label{sec:memstorage}

A memristor can be used as a digital memory of at least one bit.
The simplest way to achieve this is to use the memristor as a switch.
If the memristor is non-volatile, we can set its memristance to the value $\Ron$ or $\Roff$ using a voltage or current pulse, and associate these resistances with the state of a bit.
We stress that non-volatility is essential for memory applications because a volatile memristor, i.e., one that drifts back to a high resistance state autonomously, would require a permanent source of current/voltage to keep it in the low resistance state.
Volatility will, in general, render memristive memory worthless in terms of energy consumption.
This is one of the reasons why volatility is engineered out of the devices meant for storage applications.

In order to illustrate the mechanism of flipping a bit, consider again the non-volatile memristor model described by Equation~\eqref{eq:memristivedyn} (i.e., $\alpha = 0$) connected to a constant voltage source $V_\text{write}$.
Solving the differential equation for $w(t)$ gives:

\begin{equation}
    w(t) = \sqrt{ a + b V_\text{write} t},
    \label{eq:setwrite}
\end{equation}

\noindent where the coefficients $a,b$ depend on the parameters, but not on the input voltage.
Thus, by controlling the sign of the input voltage $V_\text{write},$ we can switch the resistance from $\Roff$ to $\Ron$ and vice versa (the~flip of a bit).
The switching happens within a characteristic time $\tau$:

\begin{equation}
\sqrt{ a + b V_\text{write} \tau} = 1 \rightarrow \tau \leq \frac{1}{b V_\text{write}}.
\end{equation}

 Hence, to flip the bit, we need to apply $V_\text{write}$ for at least this amount of time.
To read the bit, we apply a voltage $V_\text{read} \ll V_\text{write}$ (and optionally for period of time shorter than $\tau$) and compute the resistance via the measurement of the resulting current.

\begin{figure}[H]
    \centering
    \scalebox{0.9}[0.9]{
    \includegraphics{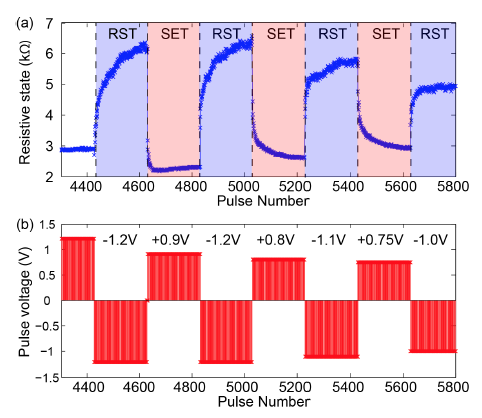}}
    \caption{$RESET$ and 
    $SET$ for TiO$_2$ with pulse within the hundred of microseconds, reproduced with permission from~\citep{Mostafa2015}. The subfigure (a) Is the evolution of 
resistance as a function of the voltage, which is shown in subfigure (b).}
    \label{fig:resetset}
\end{figure}

\textls[-15]{Equation~\eqref{eq:setwrite} applies only to the idealized memristor described by the model in \mbox{Equations~\eqref{eq:memristivedyn} and \eqref{eq:memristivestate}.}}
This model might not be valid for real devices which will show a different dynamic response to input voltages or currents.
However, the idea of controlling and reading a memristor bit with pulsed inputs remains the same.
Figure~\ref{fig:resetset} shows the response of a real TiO$_2$ 
memristor to write voltages (SET and~RESET).

The fact that it is possible to write and read the state via signal pulses allows for advantageous scaling of power consumption and bit density~(see \citep{Linn2016} for details). 
As we discuss below, it is possible to use crossbar arrays with memristors to increase the density of components.
The density of components in crossbar arrays scales as $\frac{1}{4\ell^2}$, where $\ell$ is set by the length scale of optical lithography (a few nanometers).
The reported state of the art as we write this article is roughly {7} {GB/mm$^2$}, and with writing currents of {0.1} {nA}~\citep{Shuang2018}.

Another challenge for storage based applications, besides volatility, is device variability, e.g.,~the~variation of properties like the switching time $\tau$ for devices built under similar conditions.
Current research in oxides focuses on these variability aspects, how to standardize the production of memristors, and the optimization of properties like the switching and retention time, and the durability of each singular device.
The status of the technology for memory storage for different type of devices is presented in Table~\ref{tab:techlevel}.

\begin{table}[H]
    \caption{Characteristics of various storage devices (for details see~\citep{Meena2014}).
The table compares to memristor technology to competitors like Phase-Change Materials (PCM) and more standard devices based on Spin-Transfer Torque (STT), dynamic random-access memory (DRam), Flash and hard disk~(HD)}.
    \label{tab:techlevel}
    \centering
    \begin{tabular}{ccccccc}
    \toprule
                                   & \textbf{Memristor} & \textbf{PCM} & \textbf{STT-RAM} & \textbf{DRAM} & \textbf{Flash} & \textbf{HD} \\
                                   \midrule
         Chip area per bit ($F^2$) & 4 & ~10 & 14-64 
          & 6--8 & 4--8 & n/a   \\
         Energy per bit (pJ) &  0.1--3 & $10^{1-2}$ & 0.1--1 & $10^0$ & $10^3$ & $10^4$ \\
         Read time (ns) & <10 & 20--70 & 10--30 & 10--50 & ~$10^{4-5}$ 
         & ~$10^4$ \\
         Write time (ns) & 20--30 & ~$10^{1-2}$  & ~$10^{1-2}$ & $10^1$ &  $10^{5}$ & ${10^6}$ \\
         Retention (years) 
         & 10 & 10 & $10^{-1}$ & $10^{-5}$ &  10 & 10 \\
         Cycles endurance & $10^{12}$ & $10^{7-8}$ & $10^{15}$ & $10^{17}$ & $10^{5-8}$ & $10^{15}$ \\
         3D capability & yes & no & no & no & yes & n/a\\
         \bottomrule
    \end{tabular}

\end{table}

\subsection{Crossbar Arrays}
\label{sec:crossbar}

In this section, we briefly review the crossbar array architecture used in memristor based storage and its application in artificial neural networks.

\begin{figure}[H]
    \centering
    \includegraphics[scale=0.3]{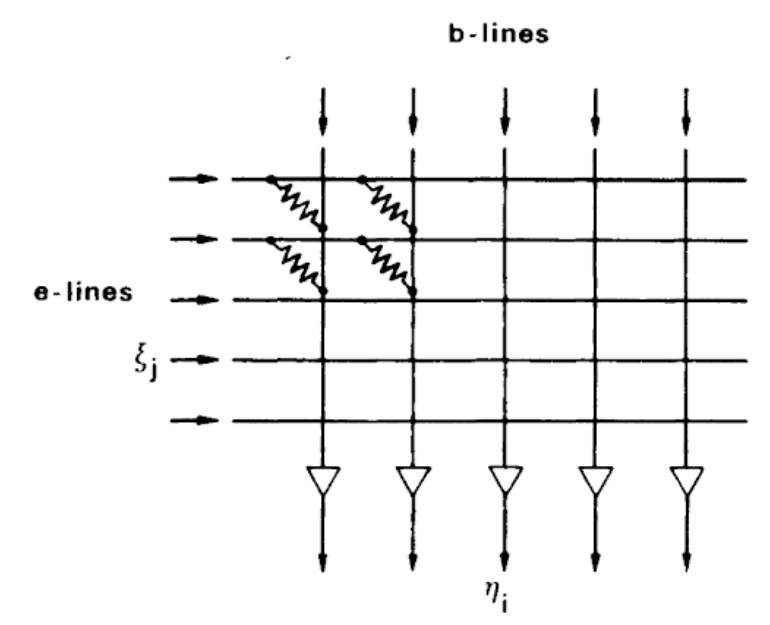}
    \caption{Learning matrix, introduced by Steinbuch~\citep{Steinbuch1961}, reproduced with permission from~\citep{Kohonen1989}.}
    \label{fig:steinbuch}
\end{figure}

Crossbar arrays are based on the architecture depicted in Figure~\ref{fig:steinbuch}.
The figure shows an array composed of horizontal ({e-lines}) and vertical ({b-lines}) lines that are initially electrically isolated from each other.
A 2-terminal component, e.g., a memristor, is connected across each pair of vertical and horizontal lines.

In order to operate crossbar arrays, a voltage $\xi_j$ is applied to the $j$-th e-line, and an another voltage $\eta_i$ to the $i$-th b-line.
A memristance $M_{j,i}$, placed across $j$-th e-line and the $i$-th b-line, is controlled by the voltage $\xi_j-\eta_i$.
This arrangement allows for simple indexing of the memristances, and is the mechanism behind a Content-Addressable-Memory (CAM), which is used in crossbar arrays.
The idea of this construction dates back to Steinbuch's ``Die Lernmatrix'' (the learning matrix)~\citep{Steinbuch1961,Steinbuch1964}.

Crossbar arrays can be used for matrix-vector multiplication using the voltages $\bset{\xi_j}$ as inputs and the voltages $\bset{\eta_i}$ as outputs.
For a resistance independent of the input voltage or current, the~relation between them is given by $\vec{\eta} = A \vec{\xi}$, where the elements of the matrix $A$ are:

\begin{equation}
    A_{ij} = \frac{ M_{ij}^{-1} }{ \tilde{R}_i^{-1} + \sum_{s=1} M_{is}^{-1} }.
\end{equation}

 $\tilde{R}_i$ are the resistances on the output b-lines $\vec{\eta}$, and $M_{ij}$ is the memristance of the memristor between the $i$-th b-line and $j$-th e-line.
An algorithm for setting the (mem)resistances given a matrix can be found in~\citep{Xia2016}.
To apply this method with memristors, the voltages differences need to be small/short, to avoid changing their memristance, or all memristors need to be in one of their limiting states $\Ron$ or~$\Roff$.

The component density of crossbar arrays can be increased by stacking various layers of crossbar arrays on top of each other~\citep{Strukov2009,Li2017}.
The stacking of $L$ crossbar layers scales the density of components by a factor $L$, i.e., a theoretical scaling of $\frac{L}{4\ell^2}$.
In multilayered arrays, memristors are controlled using the corresponding horizontal and vertical lines of each layer.

\begin{figure}[H]
    \centering
    \scalebox{0.9}[0.9]{
    \includegraphics[width=\textwidth,trim=0cm 15cm 0cm 0cm,clip]{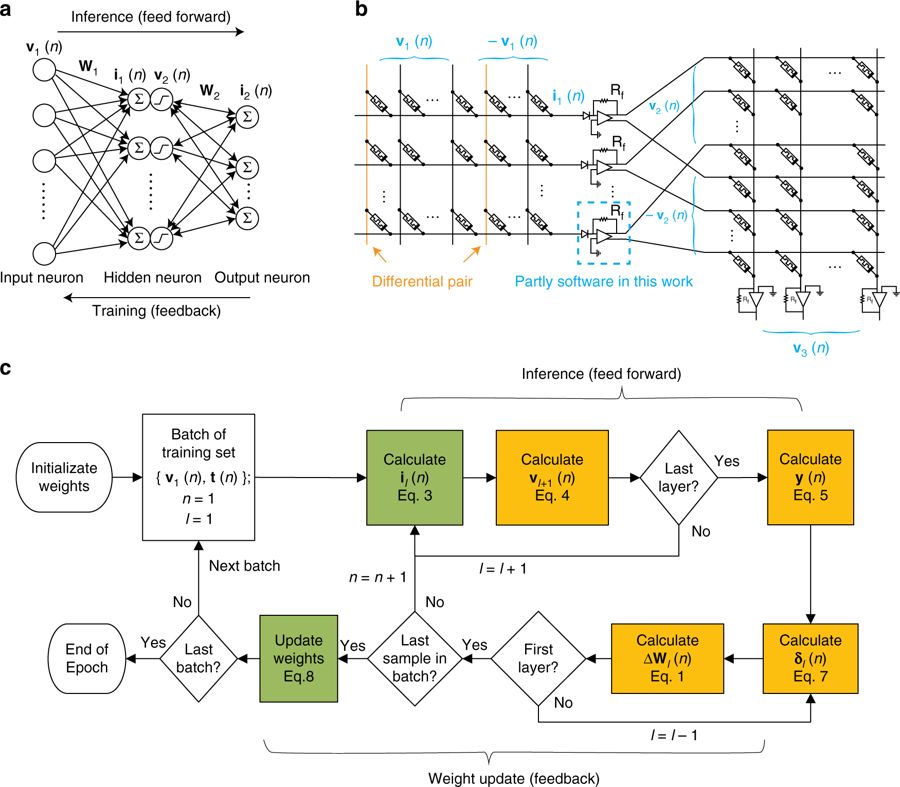}}
    \caption{Memristors used for  
    synaptic weights in artificial neural networks (ANNs), reproduced with permission from Figure  2 of~\citep{Li2018}. Subfigure (a): memristive neural network. In Subfigure (b) we introduce the crossbar implementation of the memory using memristors.}
    \label{fig:annweights}
\end{figure}

Memristive crossbar arrays can be used to encode the synaptic weights of feed-forward or recurrent artificial neural networks (ANNs)~\citep{Li2018}.
In ANNs, the input to neurons in a given layer (post-synaptic) is computed as the multiplication of the outputs of neurons in the previous layer (pre-synaptic) and the matrix of synaptic weights connecting the two layers (Figure~\ref{fig:annweights}).
Then, the~multiplication is carried over using the multiplication algorithm previously described.
The output of pre-synaptic neurons is encoded in the voltages of e-lines of the crossbar array, and the input to post-synaptic neurons is decoded from the currents on the b-lines.
Applications go beyond this direct implementation of the multiplication algorithm.
For example, in~\citep{Li2018}, the synaptic weights are encoded as the difference of the conductance between two memristors.
Similar ideas have been exploited to design other computational models based on
stateful logic~\citep{Itoh2014} and differential pair synapses (called ``kT-RAM'')~\citep{ktram}.

The~potential benefits of utilizing memristor based ANNs are speed and energy efficiency.
The~computation and storage use the same location in the network, and analog inputs are directly fed to the neurons.
This architecture minimizes the reading-writing costs incurred by the conventional von Neumann architecture, and potentially the energy losses of analog-to-digital conversion.

\subsection{Synaptic Plasticity}
\label{sec:synapplast}
Synaptic plasticity can be broadly defined as the modification of the synaptic conductance as a function of the activity of the neurons connected to it.
This definition rules out autonomous plasticity, which is the slow synaptic conductance decay of inactive synapses (volatility).
Autonomous plasticity is fundamental for data processing applications and will be considered in Section~\ref{sec:memdataproc}.

Non-autonomous synaptic plasticity can be classified in several types, e.g., spike-rate-dependent, spike-timing-dependent, short-term, long-term, etc.
In the study of synaptic plasticity of the human cortex, Hebbian or Anti-Hebbian (related to the simultaneous firing of two neurons)~\citep{Hebb1949,Kohonen1989,Koch9725} are often the underpinning learning mechanisms.
Our aim here is not to describe the types of plasticity, their biological origin, or how difficult it is to isolate each class in biological and experimental systems.
Our intention is but to illustrate how memristors, used as memories, can be used to implement the change in weights based on pre- and/or post-synaptic activity.

For an overview of the different types of synaptic plasticity and references to further reading, we refer the reader to the book chapter by~La Barbera and Alibart~\citep{LaBarbera2017}.

Synaptic plasticity is modeled by choosing the plasticity inducing variables $\bm{x} \in X$ (e.g., relative arrival times of spikes, relative neural activity, etc.) and a mapping from these to the change of the synaptic weight

\begin{equation}
  f_X: X \rightarrow W_{\Delta} \subset \mathbb{R}.
\end{equation}

 This mapping is based on biological models, or simplified adaptation mechanisms.
For example in spike-timing-dependent Plasticity (STDP), the inducing variable is the relative timing of two or more activity spikes in the connected neurons.
Plasticity is then defined with a mapping  $f_{\text{STDP}}: T^{n-1} \rightarrow W_{\Delta}$, taking the relative timing of $n$ spikes $\in T^{n-1}$ (typically 2 or 3) to a weight change $ \in W_{\Delta}$ (usually represented as relative change).
In spike-rate-dependent Plasticity, we replace the timing domain with a relative rate domain.

In order to implement plasticity in memristors as synaptic weights, we need a writing voltage that represents the synaptic change.
That is, we need a further mapping

\begin{equation}
  f_V: W_{\Delta} \rightarrow V,
\end{equation}

\noindent where $V$ is the set of valid writing voltages.
The composed mapping $ f_V \circ f_X: X \rightarrow V$ gives the final implementation of synaptic plasticity.
The mapping $f_V$ depends on all the characteristics of the technology used, e.g., the physical mechanisms of memristance (see Appendix~\ref{sec:appphys}), the neural network architecture (e.g., crossbar array), the controlling electronics, etc.
Obtaining the function $f_V$ in the mapping above is the main challenge in synaptic plasticity applications, and thus requires considerable effort.
A survey of complete and partial implementations of synaptic plasticity in nanoscale devices is summarized in~\citep{LaBarbera2017} (Section 4.1). 
For example,~Reference \citep{Ielmini2018} uses STDP to implement unsupervised learning with resistive synapses.

STDP receives a lot of attention in the neuromorphic field as exemplified by the latter reference and the review of~Serrano-Gotarredona et al., on which we base the following sentences.
The reader interested in hardware implementations of STDP should consult that resource.
STDP is among the most developed memristors implementation of in silico plasticity, it can be implemented in very large and very dense arrays of memristors without global synchronization, and learning occurs online in a single integrated phase (as opposed to offline learning).
The impact of the dynamical model of the memristor has been studied in the implementation of STDP, and the learning rules can be adapted to the different behaviors.
As in most application of memristors as non-volatile storage of (synaptic) weights, it suffers from the intrinsic variability of the units, which more general neuromorphic circuits are able to exploit~\citep{Payvand2018}.

\section{Memristors for Data Processing}
\label{sec:memdataproc}
Device variability (Section~\ref{sec:memstorage}) and volatility (Section~\ref{sec:synapplast}) were mentioned as current challenges for most applications based on memristive memories.
This is in contrast with biological systems, which are not built in clean rooms, and it is hard to believe that evolution exploits ideal systems in reference to its own architecture design.
Biological systems perform despite noise, nonlinearity, variability, lack of robustness, and volatility.
Whether these ingredients hinder the performance of biological systems or are actually a building block for it, it is still unknown.
Sometimes they are avoided, not because it would have an undesired effect in practice, but simply because its effect cannot be easily modeled or studied (e.g., we have but a few tools to deal with nonlinear systems intrinsically, beyond the iteration of linear methods).
Hence, there are no reasons to believe that eliminating naturally occurring properties is the path to success in achieving artificial systems that perform comparably to biological~ones.

This section overviews some applications that embrace device volatility~\citep{Carbajal2015}, nonlinear transients, and variability~\citep{Eliasmith2002} (Section 7.5), 
to implement learning methods with memristors.
To put these methods within an unifying framework, we first review the concept of analog computation, and discuss its relation to the physical substrate on which it is implemented.

\subsection{Analog Computation}
\label{sec:analogc}
In order to pin down the concept of analog computation, we compare analog and digital computers, assuming that the latter is familiar to most readers.
The principal distinction between analog and digital computers is that digital operates on discrete representations in discrete steps, while analog operates on continuous representations, i.e., discrete vs. continuous computation~(refer to \citep{MacLennan2012} for a complete discussion and historical overview)~\citep{MacLennan2014, MacLennan2017}.

\begin{figure}[H]
  \centering
%
%
%
\includegraphics[width=10cm]{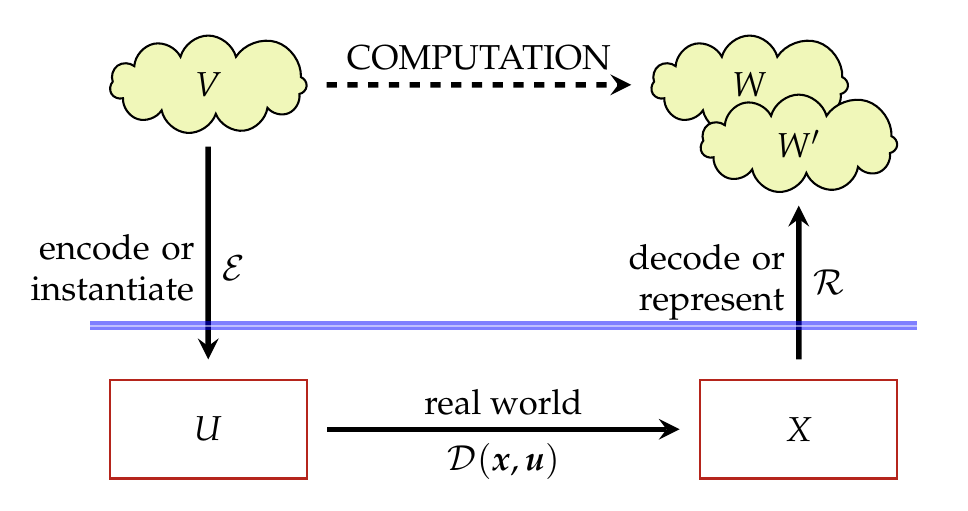}
  \caption{
    Computing with dynamical systems adapted from~\citep{Horsman2014}.
    Conceptual depiction of the sets and transformations involved in a typical computation using a dynamical system $\mathcal{D}$.
    In this case, the computation is defined by its action on the input-output set $V$, $W$.
    The inputs $\bm{u}(t)\in U$ to $\mathcal{D}$ are encoded or instantiated versions of $V$ through the transformation $\mathcal{E}$.
    The output of the dynamical system $\bm{x}(t) \in X$ is decoded or represented back into the set $W^\prime$ via the readout transformation $\mathcal{R}$.
    When $W$ and $W^\prime$ are similar, the composed transformation $\mathcal{R}\circ \mathcal{D}\circ \mathcal{E}$ is a proxy for the sought computation.
    }
    \label{fig:encode_decode_AC}
\end{figure}

In all kinds of computation, the abstract mathematical structure of the problem and the algorithm are instantiated in the states and processes of the physical system used as computer~\citep{Horsman2014}.
For~example, in the current digital computer, computation is carried out using strings of symbols that bare no physical relationship to the quantities of interest, while, in analog computers, the latter are proportional to the physical quantities used in the computation.
Figure~\ref{fig:encode_decode_AC} illustrates the relation between representation of the problem in the designers mind and instantiation in the computers states.

We can decompose computation in three stages: (i) encoding, (ii) processing, (iii) decoding.
Programming a computer implies first encoding our input to the processing unit and then decoding the output to obtain the result of the computation.
These three stages require an advanced understanding of the behavior of the physical substrate, how it reacts to inputs and how it transforms its states.
This~is true if the computer is meant to implement an universal model of computation, or if it is specialized hardware optimized to solve a particular subclass of problems.

The encoding and decoding maps relate the states of the computer to our understanding of the problem, and their definition is a recurring challenge in the design of computers.
We mentioned this difficulty when building memristive synaptic arrays with plasticity (Section~\ref{sec:synapplast}); data representation in biological systems is still an open research field, and natural systems tend to smear out our pristine categorizations of encodings.
There is yet another difficulty to attend to when defining encoding and decoding maps.
To avoid confounding, the class of maps needs to be restricted because ill-defined maps (e.g., encryption) will complicate computation, while too sophisticated maps could render the contribution of the computing device negligible.
The former will deteriorate the computing performance, while the latter is just bad design.
In other words, the problem needs to be specified using the ``language" of the computing device.
Using the wrong language increases the difficulty of the problem, and consequently decreases performance.
To understand the language of the device, we need the equivalent of Shannon's analysis of the differential analyzer~\citep{Shannon1941}.
The encoding-decoding pair is also linked to the ``natural basis of computation''~\citep{Borghetti2010} of a device, which refers to the description of the device behavior suited for the computation purposes.

The success of digital computers is in part given by the efficiency to instantiate and process a universal model of computation able to solve all kinds of computation problems, as conjectured by the Church--Turing(--Deutch) thesis~\citep{Deutsch}.
This is achieved by a precise control of each step of the computation and the way the computer transforms (or evolves) its states.
Another advantage of modern digital computers over analog prototypes is the very high precision they provide for the instantiation and solution of a problem's quantities.
This high precision, however, is unnecessary in many engineering applications, in which the input data are known to only a few digits, the equations may be approximate or derived from experiments, and the results are not sensitive to round-off errors~(see \citep{Moreau2018} for an overview), 
Therefore, research into specialized hardware (analog of digital) is a worthy activity.

Since analog computers escape the frame of relevance of the Church--Turing thesis, it has been argued that they can be more powerful than digital computers \citep{MacLennan2012} (Section ``Analog Computation and the Turing Limit''). 
Besides this benefit, it is worth exploring the efficiency of analog computers to solve subclasses of problems, i.e.,  specialized analog hardware, and to understand their pervasiveness in natural systems, perhaps linked to the precision attained by systems built from many imprecise cheap modules.
This is not a simple task, since many of these analog computers outsource some of the process control present in digital computers to the natural dynamics of the physical substrate, elevating the bar on the level of understanding required to build and program them.
It is possible to achieve a significant speedup in computational time, usually at the expense of other quantities; for~instance~\citep{Traversa2015a} reported an speedup for nondeterministic polynomial time (NP)-Complete problems at an exponential cost in energy.

As an example of an analog computer, let us consider a hydrostatic polynomial root finder proposed by~Levi~\citep{LEVISiam} .
Consider the following polynomial equation:

\begin{equation}
    \sum_{i=1}^n x^i c_i  + c_0= 0.
\end{equation}

\begin{figure}[H]
    \centering
    \includegraphics[width=0.6\textwidth]{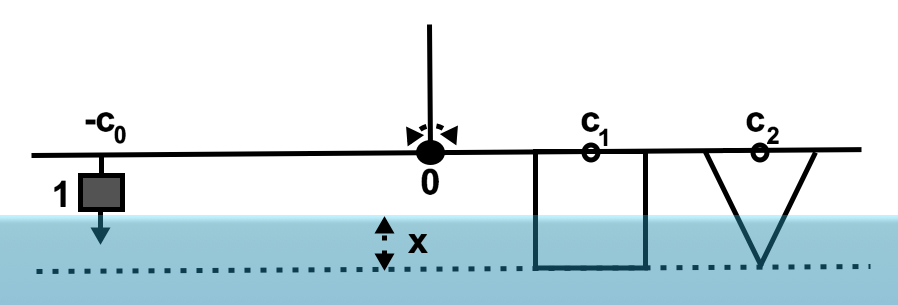}
    \caption{
    The hydrostatic root solver.
    Shapes encode the polynomial coefficients.
    A weight encodes the constant term.
    The depth at which torque is zero is a solution of the polynomial equation.
    }
    \label{fig:rootsol}
\end{figure}

 The aim is to find a real value $x$ for which the equality holds, i.e., we want one root of the polynomial $p(x) = \sum_{i=1}^n x^i c_i + c_0$.
For the sake of this illustration, let us consider the case $n=2$.
In~Figure~\ref{fig:rootsol}, we show the design of a hydrostatic root solver.
On one side of a pivoting lever, a certain shape is set to represent each term in the derivative of the polynomial (e.g., flat for the derivative of $c_1 x$, linear for the derivative of $c_2 x^2$).
On the other side of the lever, we set a weight for the constant term.
When the device is submerged into the water holding the pivoting point, the depth at which the device equilibrates (i.e., becomes horizontal) is a solution of the polynomial equation.

The example above illustrates the major role played by the understanding of the physical device, and how it allows us to encode and solve the specific problem we are interested in.
As mentioned before, this is common to all sort of computers; we proceed with mentioning a few.
The electronic digital computer exploits the behavior of transistors to encode the symbols of the computational model (essentially Boolean logic) it uses for computation.
Quantum annealers, considered for efficiently solving Mixed Integer quadratic Programming~\citep{axehill2008} (Section 2.7), 
use the spins of a physical system and the computation of a quantum Ising model to solve the optimization~\citep{Venegas-Andraca2018}.  
Deoxyribonucleic acid (DNA) computing exploits the self-assembly and preferential attachment of DNA strands to encode tiling systems with Wang tiles~\citep{Rothemund2004, Qian2011}.
Slime mold~\citep{Gale2015a} computation exploits the behavior of these protists to implement distributed optimization, e.g., shortest path, using the position of nutrients on the Petri dish to encode problems, and chemotatic behavior of the mold to solve them; the solution is read out from the distribution of mold cells in the Petri dish.
Ant colony optimization~\citep{Dorigo1997} is inspired by the behavior of their natural counterpart and uses digital models of pheromone dynamics for computation (see also stigmergy~\citep{Bonabeau1999}).

In the subsequent sections, we describe in some technical details the algorithms meant to realize computers using memristive systems.

\subsection{Generalized Linear Regression, Extreme Learning Machines, and Reservoir Computing}
\label{sec:reservoir}

Arguably, the most used method to relate a set of values $X$ (inputs) to another set of values $\bm{y}$ (outputs) via an algebraic relation is linear regression: $\bm{y} = X^\top \bm{\beta}$.
However, in many realistic situations, we believe that the relation between these sets of values is unlikely to be linear.
Hence, a nonlinear counterpart is needed.
The simplest way to generalized a scalar linear regression model between inputs and outputs is to apply a nonlinear transformation to the inputs to obtain a new linear model $\bm{y} = G(X)^\top \bm{\gamma}$, where only the coefficient vector $\bm{\gamma}$ (or matrix when the output is not scalar) is learned from the data.
This nonlinear method is a linear regression in a space generated by nonlinear transformations of the input (see kernel methods~\citep{Muller2001, Hofmann2008}).
In neuromorphic computation, the transformation is commonly given a particular structure by choosing a set of $N_g$ nonlinear functions and applying it to each input vector:

\begin{equation}
G(X)^\top_{ij} = g_j(\bm{x}(i)), \quad i=1,\ldots,N\; j=1,\ldots,N_g.
\end{equation}

This method is known as Extreme Learning Machines (ELM)~\citep{Huang2004ELM} and has been implemented in hardware exploiting the variability of fabricated devices to generate the nonlinear transformations~\mbox{\citep{Patil2017, Parmar2017}.}
Briefly explained, the differences among the devices generate different outputs for the same input, which are then used as the dictionary $G(X)^\top_{ij}$.

This formulation resembles two other mathematical methods that are worth mentioning: generalized Fourier series and generalized linear methods.
The relation with generalized Fourier series is made evident when the samples have a natural ordering (e.g., not i.i.d. time samples $i \leftrightarrow t_i$), then we can write the regression model as

\begin{equation}
y(t) = \sum_{k=1}^{N_g} \gamma_k g_k(\bm{x}(t)),
\end{equation}

\noindent which has the structure of a truncated generalized Fourier series (but not all the ingredients).

The resemblance with generalized linear models is made evident by considering the function $g(\bm{x}) = \ell^{-1}(\bm{x}^\top\bm{\eta})$ ($\ell$ is the link function) and $N_g = 1$; we obtain:

\begin{equation}
y_i = \gamma_1 \ell^{-1}(\bm{x}(i)^\top\bm{\eta}).
\end{equation}

\noindent However, generalized linear models require that we learn the vector of coefficients $\bm{\eta}$ from the data, rendering the problem nonlinear.
This breaks the analogy with ELM in which the $\bm{\eta}$ vector should be fixed a priori.
The analogy is somehow rescued if we are given a distribution $p(\bm{\eta})$ for the $\bm{\eta}$ vector encoding prior knowledge or beliefs about the solution to the generalized linear model.
In this case, we can build an ELM with $N_g \gg 1$

\begin{equation}
y_i = \sum_{k=1}^{N_g}\gamma_k \ell^{-1}(\bm{x}(i)^\top\bm{\eta}_k) \approx \int_{H} \ell^{-1}(\bm{x}(i)^\top\bm{\eta}) p(\bm{\eta})\ud\bm{\eta},
\end{equation}

\noindent where the set of $\bset{\bm{\eta}_k}$ coefficient vectors are sampled from the prior distribution, expecting that model averaging~\citep{Hoeting1999} will approximate the generalized linear model.

Summarizing, ELM uses a linear combination of a predefined dictionary of functions to approximate input-output relations.
The next step of generalization is Reservoir Computing (RC)~\citep{Jaeger01,Maass2002,Verstraeten09}, in which the $N_g$ functions are the solution of a differential or difference equations using the data as inputs, e.g.,
\begin{align}
\bm{u}(t) &= \mathcal{E}(\bm{x})(t) \; \in \mathbb{R}^{\dim\bm{u}},\\
\mathcal{D}_{\bm{\lambda}} \bm{q} &= B \bm{u}(t), \; B \in \mathbb{R}^{\dim\bm{q}\, \times\, \dim\bm{u}},\\
\bm{g} & = H\bm{q}\,, \; H \in \mathbb{R}^{N_g \times\, \dim\bm{q}},\\
y(t) &= \bm{\gamma}^\top \bm{g} = \sum_{k=1}^{N_g} \gamma_k g_k(t, \bm{u}(t), \bm{\lambda}), \label{eq:RCreadout}
\end{align}

\noindent where the differential or difference equations are denoted with $\mathcal{D}_{\bm{\lambda}}$ (operator notation) with $\bm{\lambda}$ a vector of parameters that includes physical properties and the boundary (initial) conditions, and $\dim{\bm{q}} \geq N_g$.
The connectivity matrices $B$ and $H$ are typically random, the latter mixes the $\dim\bm{q}$ states to obtain $N_g$ signals (these could also be nonlinear mappings).
As explained in Section~\ref{sec:analogc}, the input data is encoded by the transformation $\mathcal{E}$ (or $B \circ \mathcal{E}$) to properly drive the system.
This encoding, the operator $\mathcal{D}_{\bm{\lambda}}$, and the connectivity matrices are defined a priori, and only the coefficients $\bm{\gamma}$ combining the $\bm{g}$ functions are learned from the data, as in ELM.
These coefficient (or $\bm{\gamma}^\top H$) define the readout transformation $\mathcal{R}$ (see Figure~\ref{fig:encode_decode_AC}).

The generalization proposed by RC is made obvious with the choice of arguments for the component functions $\bset{g_k}$ in Equation~\eqref{eq:RCreadout}: they can have (i) an intrinsic dependence on time, e.g., autonomous behavior of the dynamical system; (ii) they depend on the inputs, and (iii) they depend on the properties of the dynamical system.
Property (ii) says that these functions are not fixed as in ELM, they are shaped by the data signal.
Stated like this, the problem of implementing computation with reservoirs is strongly related to a nonlinear control problem.

Reservoir computing (RC) allows for machine learning applications using natural or random dynamical systems, as opposed to carefully engineered ones.
Its capacity to exploit wildly different physical substrates for computation has been recently highlighted~\citep{Dale2018}.
The only strong requirement is that we are able to stimulate states of the system independently with signals encoding the input data, a classical example is the perceptron in a water bucket~\citep{Fernando2003}.
Hence, RC implementations using memristive networks has received considerable attention (software~[see \citep{Carbajal2015} and references therein] and hardware~\citep{Du2017}).
In these applications, the memristive system (single device or network) plays the role of the dynamical system $\mathcal{D}_{\bm{\lambda}}$ (cf. Figure~\ref{fig:encode_decode_AC}), the decoding map is a linear readout of several system states (memories, currents, voltage drops, etc.), and the encoding is usually tailored for the given application at hand.

The case of memristor based RC using the HP-memristor, Equations~\eqref{eq:memristivedyn} and \eqref{eq:memristivestate}, is fairly well understood: a differential equation to simulate the propagation of signals across the circuit and the interaction between memristors has been derived in~\citep{Caravelli2017} (see also Equation~\eqref{eq:memnet}).
In those equations, the role of the circuit topology is extremely important in the collective dynamics of the circuit and in processing the input information.
When working with RC and memristors, it is important to prevent the saturation of all devices, since a saturated memristor becomes a linear resistor that only scales the input.
That is, RC heavily relies in the nonlinear (and volatile) behavior of the dynamical system.

\subsection{Neural Engineering Framework}
\label{sec:NEF}
The Neural engineering framework (NEF)~\citep{Eliasmith2002} exploits our current understanding of neural data processing to implement desired computations.
It confines linear models to a particular class of basis functions, inspired by biologically plausible neuron models but not restricted to them.
Here, we describe this framework in the case of function representation, the structure of the framework is analogous to other representation instances (scalar, vector, etc.).
The reader is referred to the original work~\citep{Eliasmith2002} for a comprehensive description.

The framework entails the characterization of an admissible set of functions that can be represented by a population of $N$ neurons.
In particular, the functions and their domain need to be bounded: $f: (x_\text{min}, x_\text{max}) \rightarrow (f_\text{min}, f_\text{max})$.
These functions are then encoded by a population of neurons with a predefined set of encoders of the form:
\begin{align}
a_i(f(x)) &= G_i\left(I_i(f(x))\right) \label{eq:nefencode},\\
I_i(f(x)) &= \alpha_i \braket{f(x)\tilde{\phi}_i(x)} + I_i^{\text{bias}}, \label{eq:somacurrent}
\end{align}
 where $I_i$ represents the total input current to the $i$-th neuron soma.
The functions $a_i$ and $G_i$ are the tuning curves observed by neurophysiologists and the response function of the $i$-th neuron, respectively.
$G_i$ is a biologically inspired model of the firing rate, e.g., integrate-and-fire (LIF) neuron.
The encoding generators $\bset{\tilde{\phi}_i}$ (analogous to preferred directions) are defined a priori, and $\braket{f(x)\tilde{\phi}_i(x)}$ is a functional defined on the space of functions to be represented, e.g., in the original description this is the mean over $x$.
These encoders convert the function $f(x)$ into firing rates (or actual spike counts for the case of temporal encoding).

The encoding is matched with a corresponding decoding procedure that brings firing rates (spike counts) back to the function space.
The decoder takes the generic form:

\begin{equation}
\hat{f}(x) = \sum_{i}^N a_i(f(x)) \phi_i(x) \label{eq:nefdecode},
\end{equation}

\noindent where $\phi(x)$ are the unknowns of the framework.
That is, given some input $x$ and neural population's firing rates (spike counts), we can build functions of the input using the decoders $\bset{\phi_i}$.
In the original formulation, the decoders are obtained via minimization of least square errors (with regularization in the case of noisy encodings), but other methods could be used, e.g., optimal $L_2$ dictionaries~\citep{Sheriff2017}.

The framework defined in Equations~\eqref{eq:nefencode}--\eqref{eq:nefdecode} can realize arithmetic on functions of the input as well as nonlinear transformations.
It has also been used to represent linear time-independent systems with neuromorphic hardware~\citep{Corradi2014}.

Neural Engineering Framework (NEF), Reservoir Computing (RC), and Extreme Learning Machines (ELM) use the same form of decoding: the output is the scalar product of an input-independent vector with an input-dependent one.
The input-dependent vector is given by intrinsic properties of the computing device, it is the result of internal mechanisms.
The decoders are learned from data, and different decoders implement different computations (on the same input-dependent vectors).
However, RC and ELM learn a finite dimensional vector $\bm{\gamma} \in \mathbb{R}^{N_g}$ while NEF, in the functional form shown here, needs to learn $N$ infinite dimensional decoders.
The latter is mildly relaxed if the input domain is discretized with $N_x$ points, rendering the functional decoders $N_x$-dimensional vectors.
In general, it is expected that $N_g \ll N_x N$, which are the degrees of freedom of NEF, is higher that the one of RC and ELM.
Hence, NEF has the risk to transfer all the computation to the decoders making the contribution of the neural population marginal (or even an obstacle).
Extra regularity assumption on the decoders $\bset{\phi_i(x)}$ (Equation~\eqref{eq:nefdecode}) are needed to match decoders effective capacity to the capacity of the dynamic responses $\bset{g_i(x)}$ (Equation~\eqref{eq:RCreadout}).

Memristor networks can be used to implement NEF, by implementing the response functions $G_i$ of neural models. The $G_i$ functions corresponding to LIF neuron model is

\begin{equation}
G[I(z)] = \begin{cases}\frac{1}{\tau_0 - \tau_{\text{RC}}\log\left(1 - \frac{I_F}{I(z)}\right)}, & I(z) > I_F,\\
0, & \text{otherwise,}\end{cases}
\end{equation}

\noindent where $I(z)$ is given by Equation~\eqref{eq:somacurrent}.
A similar functional form can be achieved by a non-volatile memristor with parasitic capacitance, as shown in Equation~\eqref{eq:MClimit} (see next section).
However, other response functions, which might be easier to implement with memristors, can be used with NEF.
Other aspects of neural networks, such as critical behavior~\citep{Benna2016,Chialvo,GrossT} can be observed in networks of memristors~\citep{Avizienis}, although a theoretical understanding of their collective behavior is still poor for both systems~\citep{Caravelli2015,Caravelli2016,Sheldon2017}.

\subsection{Volatility: Autonomous Plasticity}
\label{sec:volatility}

\begin{figure}[H]
    \centering
    \includegraphics[scale=5]{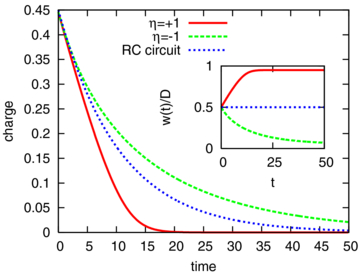}\includegraphics[scale=.6]{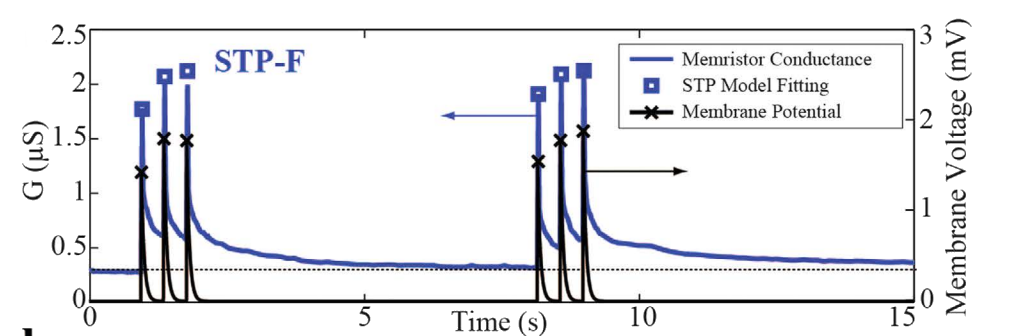}
    \caption{\textbf{Left}: 
      theoretical profile of Equation (\ref{eq:MClimit}).
    \textbf{Right}: Experimental profile of short-term plasticity with TiO$_2$.
Both figures are reproduced with permission from~\citep{Joglekar2009} and~\citep{Berdan2016}, respectively. }
    \label{fig:expstp}
\end{figure}

Volatility is a key feature when processing information with memristors (in contrast to memory applications).
RC needs volatility to avoid trivial linear input-output mappings and NEF requires it to model the forgetting behavior of neurons.
There are many physical processes that can lead to a memristive volatile device, hence the source of volatility should be discussed in the context of a given technology.
In what follows, we show how volatility can be linked with a capacitance in parallel (parasitic) to a non-volatile device.
Consider an ideal series memristor-capacitor circuit~\citep{Joglekar2009} feed with a controlled current.
The memristor is modeled with Equations~\eqref{eq:memristivedyn}, with $\alpha=0$, Kirchhoff's voltage law for this circuit gives:

\begin{align}
    R(w) \overbrace{\tder{q}{t}}^{I(t)} &=-\frac{1}{C} \overbrace{q(t)}^{\integ{I(t)}{t}} \rightarrow \tder{q}{t}(t)=-\frac{q(t)}{R(q(t))C}, \\
    R(q(t)) &= \Ron \left(1 + \frac{q(t)}{\beta}\right) + \Roff,
\end{align}

\noindent from which we obtain a limiting solution for $t \gg 1$
\begin{equation}
    q(t) = \frac{\beta}{\xi} W\left(\frac{\xi}{\beta} e^{-\frac{t}{\Ron C}} \, e^{-\frac{c_1}{\beta \Ron}}\right),
   \label{eq:MClimit}
\end{equation}

\noindent where $W$ is the product-log (Lambert) function~\citep{Veberic2010}, $c_1$ is an integration constant, and as before $\xi = \frac{\Roff-\Ron}{\Ron}$.
Equation~\eqref{eq:MClimit} shows that, for large times, the system has a typical exponential $RC$ circuit decay, which is shown in Figure~\ref{fig:expstp} (left).
This behavior is observed in experiments~\citep{Berdan2016} where after an external stimuli an exponential-like decay is observed (see Figure~\ref{fig:expstp} (right)).

The limit of a normal RC circuit can be obtained from the above in the limit $\beta\rightarrow \infty$, in which $\lim_{\beta\rightarrow \infty}\beta W(\beta^{-1} G(t))=G(t)$.
The response of a memristor-capacitor and an RC circuit differs, with the first obtaining a longer retention than the first.
In addition, the Lambert function $W$ has several properties of the logarithm function, and thus the response is similar to the one suggested in Equation~\eqref{eq:MClimit}.
In~the case $\alpha\neq 0$ for the model we described here, one has that  $w(t)=e^{t\alpha}\left(c_0+\frac{1}{\beta}\int_0^t d\tau e^{-t\tau} I(t)\right)$.
Thus, the correspondence between the parameter $w$ and the charge is not exact.
In addition, the parameter $\alpha \equiv \text{constant}$ is only an approximation.
The conductance of memristive devices decays when there is no input, and the rate of decay depends on the state of the memristor.
This is compatible with a state dependent parameter $\alpha$, rather than a constant~(see Figure 1 of \citep{Ohno2011}).
A survey of recent hardware designs for temporal memory is provided in~\citep{SurveyTempNet}.

\subsection{Basis of Computation}

As mentioned before, to design computers with memristors, we need to understand and harness their natural computational power.
This is no simple task in general, but for networks of current-controlled memristors linear in an internal parameter, we can write down the differential equation describing the evolution of the memory states and obeying Kirchhoff voltage and current laws~\citep{Caravelli2017}:
\begin{equation}
    \tder{\vec{w}}{t}(t)=\alpha \vec{w}(t)-\frac{1}{\beta} (I+\frac{\Roff-\Ron}{\Ron} \Omega W)^{-1} \Omega \vec S(t),
    \label{eq:memnet}
\end{equation}

\noindent where $\alpha$ and $\beta$ are the parameters in Equations~\eqref{eq:memristivedyn}--\eqref{eq:memristivestate}, $\Omega$ is a projector operator which depends on the circuit topology, $W_{ij} (t)=\delta_{ij} w_i(t)$ and $\vec S(t)$ is vector of applied voltages.
For arbitrary memristor components, the generalization of Equation~\eqref{eq:memnet} is not known.
In the approximation $\Roff= p \Ron$, with $p$ of order one, the equation above can be recast in the form of a (constrained) gradient descent~\citep{Caravelli2018}, which is reminiscent of the fact that the dynamics of a purely memristive circuit has an approximate Lyapunov function~\citep{Caravelli2018b,Caravell2018mfgen}.
In the simplified setting of purely memristive circuit without any other components, it can be shown that these circuits execute Quadratically Unconstrained Binary Optimization~\citep{Boros2007}.
This idea is in general not recent, and it can be traced back to Hopfield~\citep{HopMemr1,HopMemr2,HopMemr3} for continuous neurons.
It is also not the only alternative approach to Hopfield networks using memristors~\citep{Hopmemror1,Hopmemror2,Hopmemror3,Hopmemror4}, as we will also  discuss it later.

\section{Memristive Galore!}
\label{sec:memgalore}
\vspace{-6pt}
\subsection{Memristive Computing}
\label{sec:memcomp}
In this section, we review some works developing computation algorithms using networks of memristors and external control hardware based in crossbar arrays and FPGA.

\begin{figure}[H]
    \centering
    \includegraphics{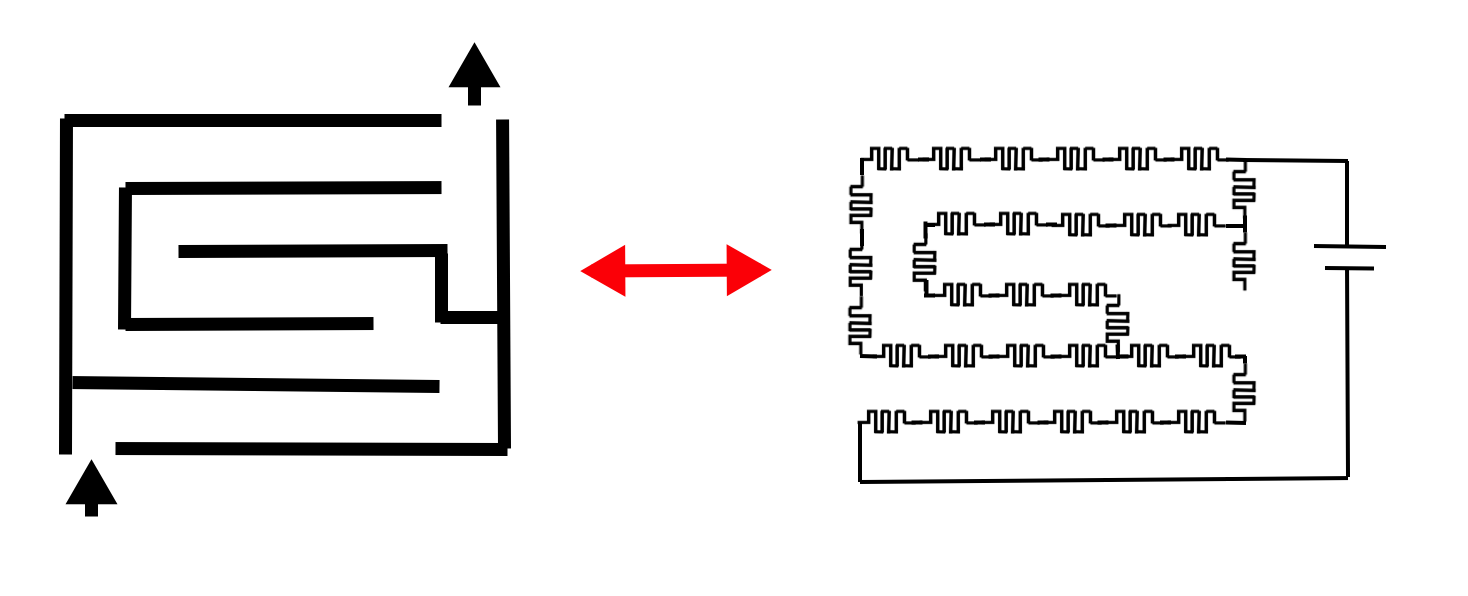}
    \caption{The maze-memristor mapping suggested in~\citep{Pershin2011}.}
    \label{fig:maze}
\end{figure}

In~\citep{Pershin2011,Pershin2013}, it has been shown that memristive circuits can be used to solve mazes: connecting the entrance and exit of a maze, the memristive circuit as in Figure~\ref{fig:maze} will re-organize (when controlled in direct current (DC)) to allow 
the majority of the current to flow along the shortest path.
Although this phenomenon already occurs with regular resistances (with a diode in parallel), it is enhanced with memristors.
Memristors outside the shortest path go to their OFF state (high resistance) and the current difference (the contrast) to the active shortest path is augmented.
This example shows that the wiring between memristors and the asymptotic resistance values are deeply connected; this fact is reminiscent of the ant-colony optimization algorithms~\citep{Dorigo1997}, molecular computation~\citep{MolCop}, and other cellular automata models~\citep{Adamatzky1996}.

Ideas along these lines can be pushed further in order to explore memristors as complex adaptive systems~\citep{prokopenko} able to self-organize with the guidance of the circuit topology and the control of external voltages.
Using Equation~\eqref{eq:memnet}, it is possible to obtain approximate solutions, for instance, of the combinatorial Markowitz problem~\citep{Caravell2018mfgen}.
Hybrid CMOS/Memristive circuits can, in principle, tackle harder problems via a combination of external control and self-organization~\citep{WilliamsSOM}.

In the literature, several models of memristor-based architectures have been proposed.
Many of these proposals are based on the attractor dynamics of volatile dissipative electronics and inspired by biological systems.
For instance, a general theory of computing architecture based on memory components (memcapacitors, meminductors and memristors~\citep{memall}) has been introduced recently in~\citep{UMM}, called Universal Memory Machines (UMM), and shown to be Turing complete.
Similarly, an architecture based on memristors which includes Anti-Hebbian and Hebbian learning (AHaH) has been proposed in~\citep{ahah,ktram} for the purpose of building logic gates and for machine learning.
In both cases of UMM and AHaH, the solutions of the problems under scrutiny are theoretically embedded in the attractors structure of the proposed dynamical systems, and have not been tested experimentally.

One way to show that a memristive system is a universal computing architecture is to break the system modularly into logic gates based on memristors, and show that the set of obtained gates is universal (which includes NOT and at least one of an AND or OR gates, as in DeMorgan's law~\citep{logicbook}).
Turing completeness follows from an infinite random access memory (the infinite tape).
Experimentally, it has been shown that it is possible to build logic gates with memristors (we mention for instance~\mbox{\citep{galelog,loggatimp}}).
An improvement upon this basic idea is to build input-output agnostic logic gates using memristors.
Any port of an agnostic gate can be used as input or output, and the remaining states of the gate will converge to the states of a logic gate, regardless of whether the binary variable is at the output or at the input of the gate.
For example, if the output of an agnostic AND gate is set to TRUE, the input variables will rearrange to be both TRUE, but, if the output is FALSE, the inputs will be re-arranged such as to contain at least one FALSE.
These are called Self-Organizing Logic Gates (SOLG) and it is suggested to use these to solve the max-SAT problem~\citep{TraversaPol,maxsat}~(see also~\citep{diventratravrev} and references therein). Similar ideas were recently proposed using nanoscale magnetic materials rather than memristors~\citep{boospin}.

The two examples above show that memristors can be used both for analog computation, as in the case of shortest path problems, or to reproduce and extend the properties of digital logic gates.

\subsection{Natural Memristive Information Processing Systems: Squids, Plants, and Amoebae}

In recent years, and with the participation of Chua, there have been several reports re-interpreting models of natural information processing systems (neural networks, chemical signaling, etc.) in terms of memristors units.
Here, we mention three examples: giant axon of squids~\citep{Sah2014}, electrical networks of some plants~\citep{Markin2014}, and Amoeba adaptation~\citep{AmLeao}.

\begin{figure}[H]
     \centering
     \includegraphics[scale=2.2]{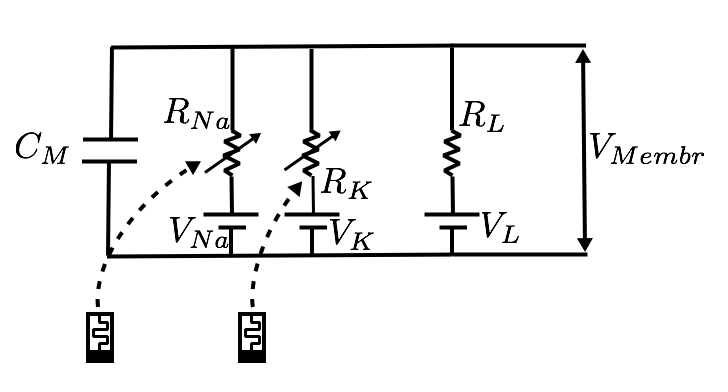}
     \caption{The Hodgkin--Huxley model, with the variable resistances of the Sodium and Potassium channels interpreted as memristors.}
     \label{fig:hhmem}
 \end{figure}

{\textbf{Squids}.} 
Giant squids are model organisms big enough that they can be analyzed in detail at the singular cell level, and for which we posses a mechanistic model of the dynamics of their axons: the Hudgkin--Huxley (HH) model.
The HH model describes the voltage at the interface between synapses and dendrites, which is regulated by the flow of calcium and potassium.
The electrical circuit associated with this model is shown in Figure~\ref{fig:hhmem}.
It entails the introduction of a nonlinear variable resistor for the calcium channel, and a linear variable resistance for the potassium channel and a capacitance~\citep{HHmodel}.
The equations of this model, when put in memristive form, are given by:
\begin{align}
i_{\text{K}}  &= \overbrace{g_{\text{K}} w_1^4}^{R_{\text{K}}^{-1}} V_{\text{K}} ,\\
i_{\text{Na}} &= \overbrace{g_{Na} w_2^3 w_3}^{R_{\text{Na}}^{-1}} V_{\text{Na}} ,\\
\tder{w_1}{t} &= \left( \text{K}_1\, V_{\text{K}} + \text{K}_2 \right) \left[e^{\,\text{K}_1\, V_{\text{K}} + \text{K}_2} - 1 \right]^{-1} \, \left( 1 - w_1 \right), \\
\tder{w_2}{t} &= \left( \text{Na}_1\, V_{\text{Na}} + \text{Na}_2 \right) \left(e^{\, \text{Na}_1\, V_{\text{Na}} + \text{Na}_2} - 1 \right)^{-1} \, \left( 1 - w_2 \right)  + \text{Na}_3\, e^{\, \text{Na}_4\, V_{\text{Na}} + \text{Na}_5} \, w_2,\\
\tder{w_3}{t} &= \text{Na}_6\, e^{\, \text{Na}_7\, V_{\text{Na}} + \text{Na}_8} \, \left( 1 - w_3 \right) - \left(e^{\, \text{Na}_1\, V_{\text{Na}} + \text{Na}_{9}} + 1 \right)^{-1} \, w_3,
\end{align}

\noindent where we see that a first order ($R_{\text{K}}$) and second order ($R_{\text{Na}}$) memristors are involved.
The parameters $\bset{\text{K}_i}$ and $\bset{\text{Na}_i}$ characterize the dynamics of the channels~(see \citep{Sah2014} for details).

The model above is a fit of the observed voltage data for the giant axon, and is useful in the analytical study of brain cell dynamics.
The proposed model allows the interpretation of synapses as circuits composed of rather nonlinear and non-ideal memristors.
That is, that memristors can be central in providing an alternative interpretation of a established model and further the understanding of biological neural information processing.

\textbf{Plants.} 
In~\citep{Markin2014}, three types of memristors models were developed and compared with the responses of some plants to periodic electrical stimulation.
The authors observed that, in the studied plants, the pinched hysteresis loop did not collapse into a line for very high frequencies, as required for ideal memristors.
To recover this non-ideal behavior, a parasitic resistor--capacitor pair was added in parallel to the ideal memristor model.
The general solution for their models is:

\begin{align}
i_m(t) &= \frac{e^{\beta t} V(t)}{\beta R_o\int_{0}^{t} h(V(x)) e^{\beta x} \ud x + A},\\
I &= i_m + i_{RC},
\end{align}
\noindent where $\beta$ is a parameter related to the time constant of the memristor, $V(t)$ is the driving periodic voltage, and $R_o h(V)$ is the memristance of a voltage-controlled memristor.
Depending on the model of the memristor considered, $h$ and the constant $A$ take different forms.
The total current $I$ is the observed magnitude, and $i_{RC}$ is the current through a series resistor--capacitor circuit in parallel with the memristor.
The study hints that memristive behavior is intrinsic to plants electrical signaling and that plant physiology could be better understood if memristors are considered as ``essential model building blocks''.

\begin{figure}[H]
    \centering
    \includegraphics[width=0.5\textwidth]{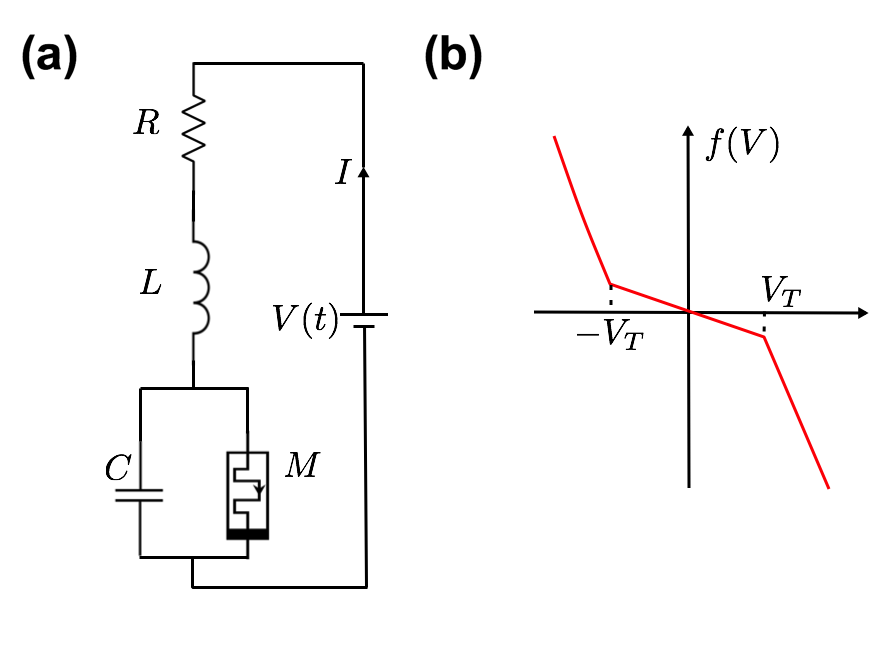}
    \caption{The amoeba memristive learning model of~\citep{AmLea}.
(\textbf{a}) the circuit with capacitance $C$ and memristor $M$ in parallel (with resistance $R(t)$), and in series to a resistance $R$ and an inductance $L$;
(\textbf{b})~the function $f(V)$ for the memristor response in voltage.}
    \label{fig:aml}
\end{figure}

\textbf{Amoeba.} 
A memristive model of amoeba adaptation was introduced in the form of the simple circuit shown in Figure~\ref{fig:aml}~\citep{AmLea,AmLeao}, and the model is simple enough that we can report it here.
Albeit the original article points to the concept of amoeba learning, we believe it is more appropriate to be addressed as a model of amoeba adaptation.
The memristor considered is a voltage-controlled memristor introduced in~\citep{WilliamsVCM}:
\begin{align}
    \tder{M}{t} &= f(V_M) \left(\theta(V_M) \theta(M - R_1) + \theta(-V_M) \theta(R_2 - M) \right),\\
    f(V) &= \frac{\beta - \alpha}{2} \left( \vert V + V_T \vert - \vert V-V_T \vert\right) -\beta V,
\end{align}
\noindent where $\theta(\cdot)$ is the Heaviside step function.

Because the inductance and the resistor in the circuit are in series, the same current $I$ flows through them.
The capacitor and the memristor are in parallel, hence their voltage drop are equal: $V_C = V_M$.
The conservation of voltage on the mesh implies $V_C + V_L + V_R = V(t)$.
We have $V_R = R I$ and $V_L = L I$.
The memristance $M(t)$ affects the voltage drop on the capacitor,
\begin{equation}
C V_C + \frac{V_C}{M(t)} = I.
\end{equation}

 We thus obtain the three coupled differential equations:
\begin{align}
 \tder{I}{t} &= -\frac{R}{L}I + \frac{V - V_C}{L}, \\
 \tder{V_C}{t} &= -\frac{1}{M C}V_C + \frac{I}{C}, \\
 \tder{M}{t} &= f(V_M) \left( \theta(V_M) \theta(M - R_1) + \theta(-V_M) \theta(R_2 - R) \right).
 \label{eq:simam}
\end{align}

\begin{figure}[H]
    \centering
    \includegraphics[width=0.6\textwidth]{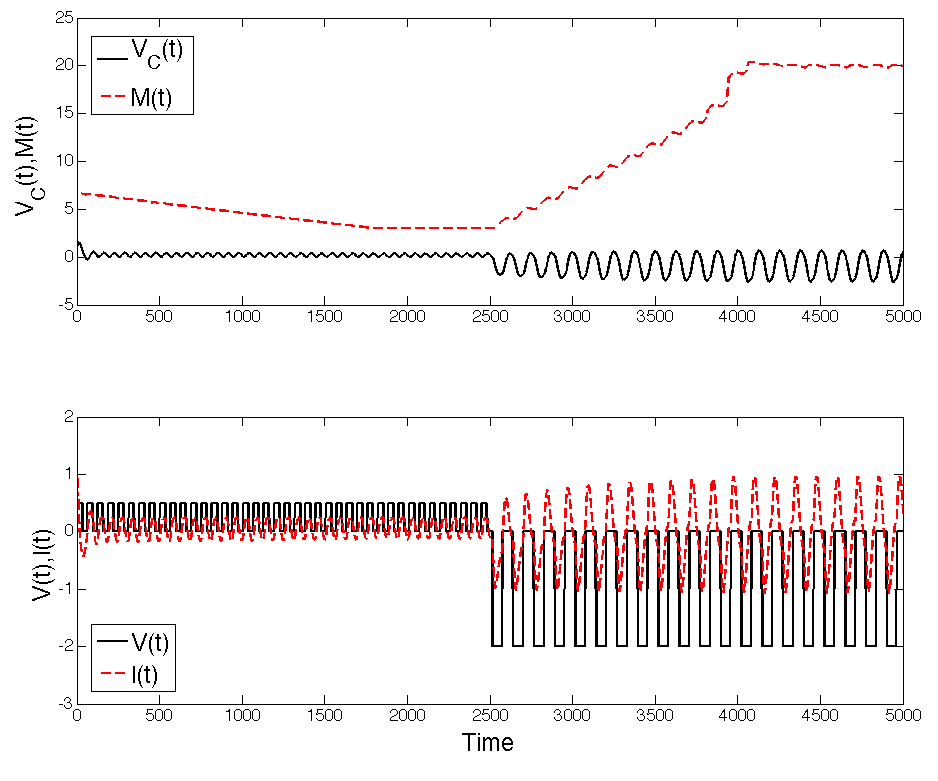}
    \caption{Simulation of Equation~\eqref{eq:simam} and adaptation of the circuit to different stimuli.
We consider the parameters $\beta=100$, $\alpha=0.1$,
$R_1=3$, $R_2=20$, $C=1$, $R=1$, $L=2$, $V_t=2.5$ as in~\citep{AmLea}.
We~stimulate the circuit with a square input $V(t)=0.5$ and frequency $\omega=10$ and then with reduce by a half the frequency, with $V(t)=-2$.
We see that the circuits ``adapts'' to the new stimulus after a transient.
Initial conditions were $I_0=1$, $V_c^0=1, R(0)=7$ and used an Euler integration scheme with step $dt=0.1$.}
    \label{fig:matam}
\end{figure}

 The stationary state requires that all time derivatives are zero.
In this state, the circuit is sensitive to new stimuli, i.e., it adapts.
For instance, in Figure~\ref{fig:matam}, we see the response of the system to new inputs, and new stationary states are obtained.
Albeit amoeba's adaptation are not as developed as in higher mammals, more general models entailing Pavloviav ``conditioning" have also been proposed in the literature using memristors~\citep{Pershin2010,Tan2017}.

\subsection{Self-Organized Critically in Networks of Memristors}

We now consider the interaction between a high number of components with memory.
A~common \textls[-15]{feature of large systems of interacting units with thresholds or discontinuous dynamics is critical behavior.
For example, self-organized criticality (SOC) is evinced when a dynamical system self-tunes into a configuration for which a qualitative change in the systems' behavior is imminent (e.g., a bifurcation).}
These critical configurations are characterized by states with power law cross-correlation functions.
One of the main motivations of SOC is the explanation of power spectra of the functional form $P\sim \omega^\alpha$, with $-1<\alpha<-3$, in physical systems and in nature~\citep{Turcotte1999}.
The~typical example is, for instance, the Gutenberg--Richter law of earthquakes, whose distribution of magnitude is Richter's law~\citep{Jensen1998,Markovic2014}.
The~current characterization of the sufficient ingredients for SOC, however, is phenomenological: a system of interacting particles or agents in which thresholds are present and whose dynamics is dominated by their mutual interaction.
SOC has been suggested to be the underlying mechanism in the observed critical behavior of the brain~\citep{Chialvo}.
In this case, neurons can be interpreted as thresholding functions (logical gates) and it is thus tempting to interpret the criticality of the brain as a SOC~phenomenon.

SOC can be produced using large networks of memristors: atomic switch networks are dominated by the interaction due to Kirchhoff laws~\citep{Avizienis}, and thus the observed criticality seems intuitively (power law distribution of the power spectra, for instance) to be connected to a SOC-type~\citep{Sheldon2017} phenomenon because of the rather nonlinear and threshold-like behavior of memristors.
It is however easy to observe that power law distributions in power spectra can be obtained in a rather simple way as follows~\citep{Caravelli2017}.
Consider a system of interacting memristors, whose linearized dynamics close to a fixed point obtained with DC voltage stimulation (i.e., saturated memristors: resistors) is written as:

 \begin{equation}
     \tder{\vec{w}(t)}{t}= A \vec{w}(t).
 \end{equation}
 
 The matrix $A$ is a non-trivial combination of voltage sources and projectors on the subspace of the circuit's graph~\citep{Caravelli2018}.
We divide the spectrum of $A$ in positive and negative eigenvalues, and the distribution $\rho_+(A)$ and $\rho_-(A)$.
Since $0\leq w_i(t)\leq 1$, we look at the average relaxation $\langle w(t)\rangle=\sum_{i} \frac{ w_i(t)}{N}$, which can be written as
\begin{equation}
   \langle w(t)\rangle= \frac{1}{N}\sum_i \left( \sum_j (e^{A t})_{ij} w_j^0 \right)=\frac{1}{N}\trace\left(e^{At} W_0\right)=\frac{1}{N}\trace\left(e^{\lambda^+t} \tilde W^0+e^{\lambda^-t} \tilde W^0\right).
\end{equation}

 The positive part of the spectrum will push memristors to the $w=1$ state, while the negative part to the $w=0$ state.
We can write the trace on each positive and negative state as:

\begin{equation}
    \frac{1}{N}\trace(e^{-\lambda^{-}_i t} \tilde w_i)=\int d\lambda \rho^{-}(\lambda) e^{-\lambda t} \langle w_i^0\rangle = \frac{1}{2}\int d\lambda \rho^{-}(\lambda) e^{-\lambda t},
\end{equation}

\noindent if the memristors are randomly initialized.
As it turns out, if $\rho^{-}(\lambda)$ is power law distributed, then $\langle w(t) \rangle \approx t^\gamma$.
From this, we observe that the power spectrum distribution is of the form $P(\omega)\approx \omega^{-(1-\gamma)}$, from which a ``critical state" is obtained.
It was shown in~\citep{Caravelli2017} that if the circuit is random enough, numerical simulations produce $\gamma\approx -1$.
A similar argument can be obtained for $\rho^+(\gamma)$ with the transformation $w\rightarrow 1-w$.
This result, when using networks of HP-memristors, is in line with what is experimentally observed in~\citep{Avizienis}.

The phenomenon above is not classified as SOC, since only a certain matrix $A$ is required to obtain it.
This implies that unless thresholding is present, the criticality observed is not self-tuned, but it might be due to the complex interconnections.
This is not the case if, however, the memristors themselves have voltage induced switching~\citep{Sheldon2017}.
In this case, the criticality is due to the a SOC-like phenomenon which had been already observed in random fuse networks, which is described by a percolation transition~\citep{Alava2006}.

\subsection{Memristors and CMOS}
\label{sec:experiments}
\begin{figure}[H]
    \centering
    \includegraphics[scale=1.5]{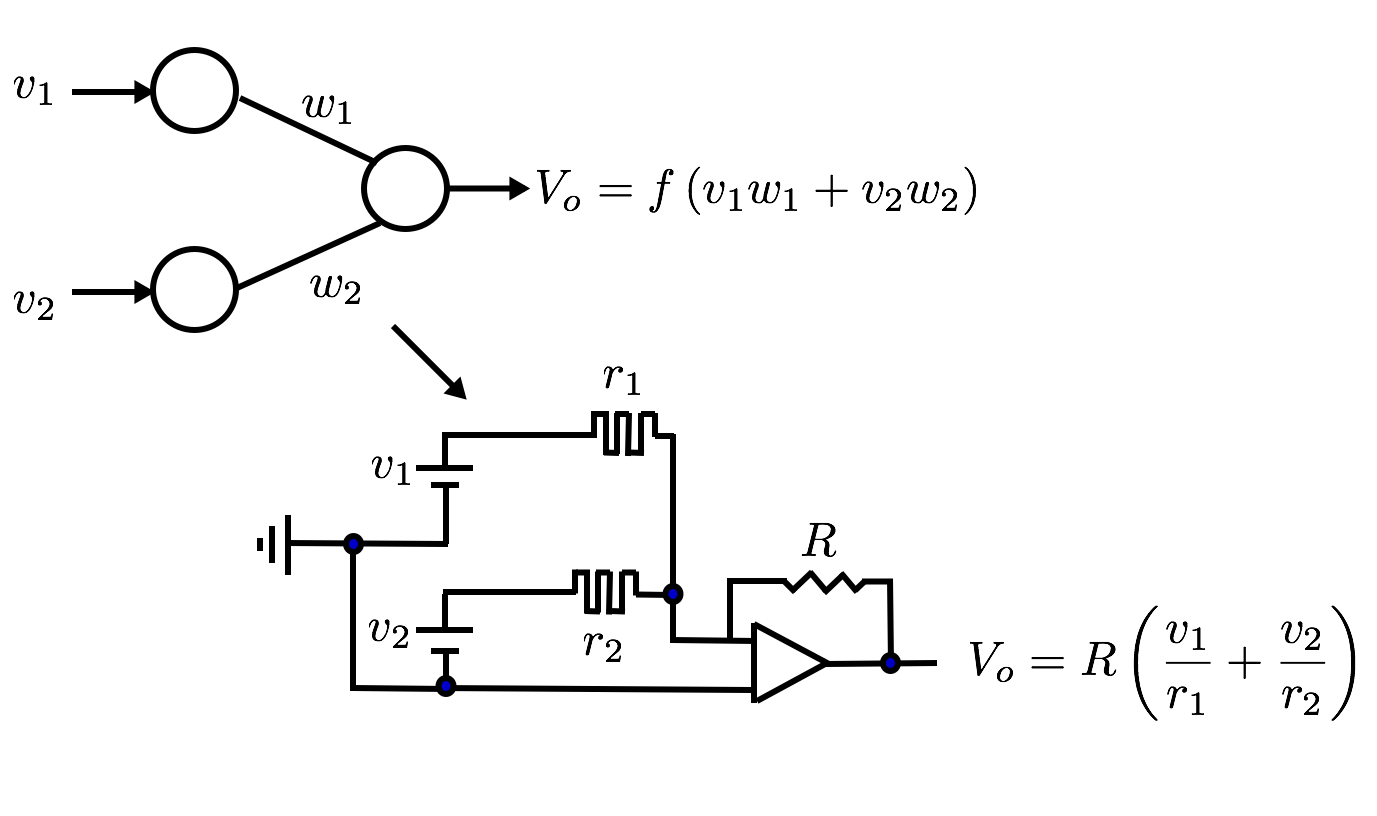}
    \caption{Memristor equivalent of a neural network with three neurons and two synapses, with an output amplifier.}
    \label{fig:nnMM}
\end{figure}

The idea of using variable resistances in order to implement learning algorithms is not new.
As we have seen, crossbar arrays were introduced already as early as 1961~\citep{Steinbuch1961}.
The idea of using instead variables resistances precedes the paper of Steinbuch by one year, and was introduced by Widrow~\citep{Widrow1960} in order to implement the Adaline algorithm explained below. 
The ``\textbf{mem}istor'', a~name extremely similar to the one of ``\textbf{memr}istor'' introduced ten years later by Chua, was a current-controlled variable resistance~\citep{Adhikari2014}.
This limited the ability to package a huge number of memistor synapses.
For~(modern) machine learning applications, however, it is necessary to implement learning rules with a large number of neurons and synapses.
As we have seen, memristors are the equivalent component for a synapse~\citep{memsynapses}.
The neuron is instead the biological equivalent of a $N$-port logic gate (threshold function).
For more general applications, a crossbar-array like packing is desirable.
There are many ways of using memristors for applications in neural networks~\citep{reviewGorm}.
For instance, the circuit proposed in Figure~\ref{fig:nnMM} provides a simple circuit which has a linear output neuron controlled by a resistance $R$, without a threshold.
The threshold can be however easily introduce via a Zener diode.
In order to understand why memristors do not have necessarily an advantage over digital implementations of neural networks, we follow the argument of~\citep{DeBenedictis2016} to understand the energy efficiency.
Using Landauer theory, the energy dissipated by a digital gate is $E_{gate}\approx -2 \log(p_{err}) k T$, where $p_{err}$ is the probability of an error.
For the analog implementation above, the only dissipation is due to Johnson--Nyquist noise in the amplifier, which is of the order $4 kT f$, with $f$ the amplifier's bandwidth and $N$ the number of synapses.
Keeping track of error and number of bit precision $L$, one reaches the conclusion that
\begin{align}
    E_{dig}&\approx 24 \log(\frac{1}{p_{err}}) \log_2^2(L) N k T,\\
    E_{memr}&\approx \frac{1}{24} \log(\frac{1}{p_{err}}) L^2 N^2 k T,
\end{align}
which scales in the number of artificial neurons in favor of digital implementations.
This surprising result thus confirms that the devil is in the details, and that not necessarily all analog implementations of digital systems are better.

The architecture of Figure~\ref{fig:nnMM}, however, does not take advantage of the scalability of crossbar arrays which we have mentioned earlier, in terms of number of neurons, and other implementations might be more energy efficient.
For instance, in~\citep{Alibart2013}, first experimental results of memristive technology for pattern classification on a 3 $\times$ 3 image matrix was studied using crossbars and TiO$_2$ memristors.
In general, learning using crossbars follows a general weight update strategy.
For regressions, current-controlled memristors can be used with a simple serial architecture~\citep{Carbajal2015} while unsupervised learning can be performed by implementing the K-means algorithm~\citep{KMeansmemr,UnsLern,Jeong2018}.

\begin{figure}[H]
    \centering
    \includegraphics[width=0.5\textwidth]{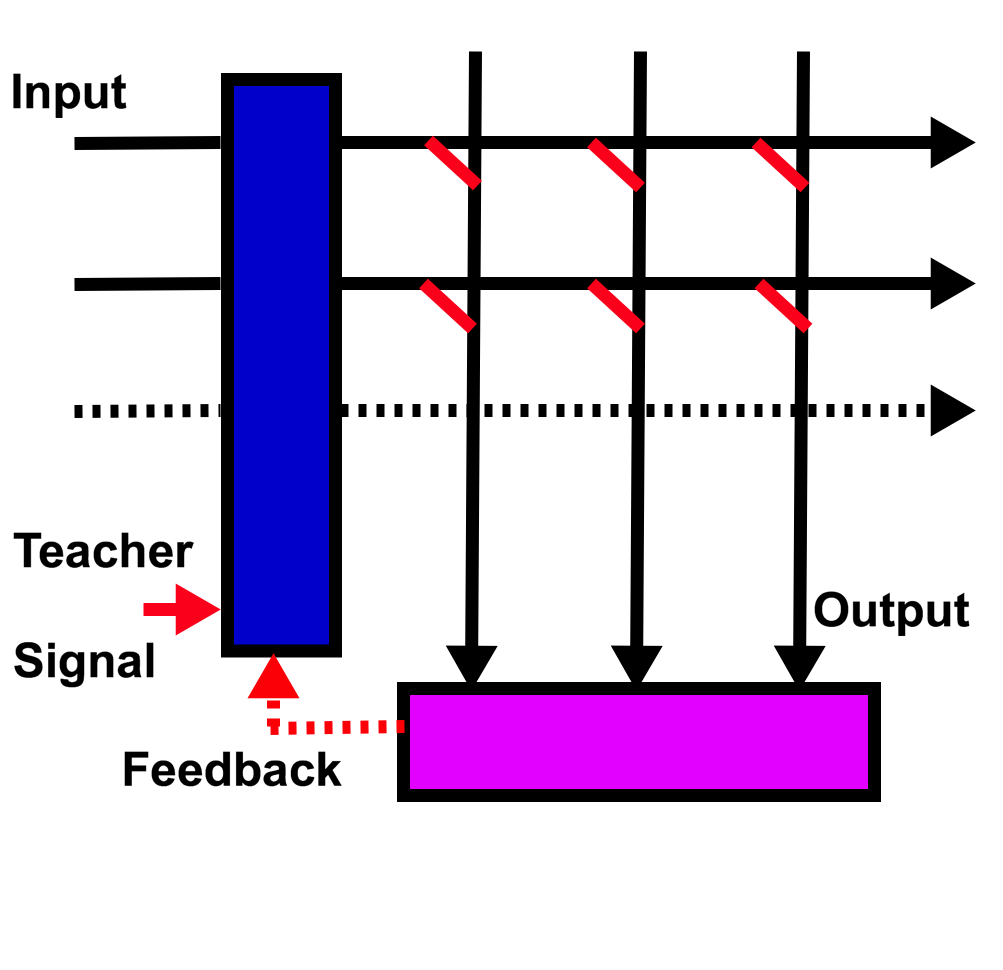}
    \vspace{-18pt}
    \caption{Feedback loop and learning in crossbar arrays.}
    \label{fig:feedbackcrossbar}
\end{figure}

Next, we focus on applications of memristor technology for Machine Learning (ML).
Algorithms like backpropagation on conventional general-purpose digital hardware (i.e., von Neumann architecture) are highly inefficient: one reason for this is the physical separation between the memory storage (RAM) of synaptic weights and the arithmetic module (CPU), which is used to compute the update rules.
The bus between CPU and RAM acts as a bottleneck, commonly called von Neumann bottleneck.
As mentioned before, the idea is thus to use memristive based technology to introduce computation and storage on the same platform.
One way to introduce learning into crossbars is by introducing feedback into the system, as in Figure~\ref{fig:feedbackcrossbar}.

Let us consider a discretized dynamics, and call $W^k$ the weight matrix in the crossbar array at time step $k$.
A general update is of the form

\begin{equation}
    W^{k+1}=W^{k}+ f(W^k, Q),
    \label{eq:wupdate}
\end{equation}

\noindent where $f(W^k)$ is a certain function of the weights and some ancillary variables $Q$ (which could be training data, inputs, outputs, etc.).
For instance, in the case of neural networks training (gradient descent type),
$f(W^k)=- \eta \nabla_{W^k} \Vert\vec t-\vec o\Vert^2,$
where $\vec t$ is the output we aim to obtain (given the inputs), and $\vec o$ is the output.
For resistive crossbars, we have seen that $\vec o=G(W^k) \vec v$ is linear in the inputs and $G$ is the conductance, while $\eta$ is a time scale parameter.
If $f(W)_{mn}=\eta x_{m}^k x_{n}^k$, where $x_{n}^k$
 is the teacher inputs (patterns) indexed by the index $k=1,\cdots, K$, then the Hebbian learning rule is called adaline algorithm~\citep{Hebb1949,Widrow1990}.
 From the point of view of circuit theory, the feedback can be introduced via CMOS-Memristor integration.
For instance, in~\citep{Soudry2015}, one way to perform online learning with an adaline algorithm using metal-oxide-semiconductor field-effect transistors (MOSFETs) has been~provided.

Models like the one just described can, for instance, be used for sparse coding~\citep{Rozell2008}.
Sparse coding can be mathematically formulated as a problem in linear algebra.
Consider a vector which describes a certain quantity of interest $\vec x$ (for instance, an image), and a dictionary which is available as $\vec \phi^i$, $i=1,\cdots,M$.
Furthermore, assume that $\phi^i$ and $\vec x$ belong to $\mathbb{R}^N$.
If $M>N$ and the vectors are independent, we can always find the coefficients $a_i$ such that $\sum_i a_i \vec \phi^i=\vec x$, and there is an infinite number of ways to do this expansion.
The goal of sparse coding is solving the following optimization~problem:
\begin{align}
\vec x&=\sum_{i=1}^M a_i \vec \phi^i, \\
\text{min}_{\vec a}& \Vert\vec a\Vert_0,\ \Vert\cdot\Vert\text{ 0-norm},
\end{align}

\noindent and which is notoriously NP-hard. The problem above can be relaxed by replacing the 0-norm with the 1-norm, and a system of differential equations for continuous neurons can be implemented in crossbar arrays.
In general, these are equations of the type
\begin{equation}
    \tder{\vec{u}(t)}{t}=F(\vec{u}(t),\vec{q}(t)),
\end{equation}
where $u_i(t)$ and $q(t)$ are some control functions and $F(\cdot)$ represents a generic continuous function. Some details are provided in Appendix \ref{sec:scod}. The variables $u_i(t)$ are intended as the memory elements in a memristor:
in a crossbar array system, the equations above have been implemented by combining a field programmable gate array (FPGA) with a crossbar in~\citep{Sheridan2017}, where using a threshold function $T_\lambda(x)=x$ if $x>\lambda$ and zero otherwise.
Another update rule used in experiments is Sanger's update rule~\citep{Sanger1989}, defined as

\begin{equation}
    W^{k+1}=W^k+2 \eta \vec{o}^t ( \vec{x}-(2W^k -I)\vec{o}),
\end{equation}

\noindent where $\vec{o}$ is the output of the crossbar array, and $\vec{x}$ the input, which is used in~\citep{Choi2015} in order to perform PCA, again with the use of FPGA.

Recent advances in Deep Neural Network implementations using memristor technology (PCM) show remarkable efficiency in terms of energy consumption for a given constant accuracy, with gains up to two order of magnitudes when compared to graphics processing unit (GPU)
implementations~\citep{Hopmemror3}.
The architecture used therein is not dissimilar from the one described in this section using a combination of CMOS and crossbar arrays.
This is another example of how crossbar arrays are amenable to efficient matrix multiplication, which renders them competitive to other GPU implementations~\citep{Hopmemror4,HopMemr1,HopMemr2}.

We have already discussed reservoir computing (RC)~\citep{Carbajal2015,Du2017} as another way of using memristors within the framework of machine learning whilst taking advantage of their temporal dynamics (see Section~\ref{sec:reservoir}).
In~\citep{Du2017}, RC was implemented with a memristor reservoir and one layer output with \mbox{32 $\times$ 32} crossbar of memristors and trained using an external FPGA.
The model was then used successfully to classify the MNIST dataset (5 $\times$ 4 images) as a proof of principle application.
The advantage of using RC is that it can be implemented both for online learning and for classification in a teacher--signal framework, and with relatively little computational effort.

In conclusions, CMOS provides an advantage for controlling memristive circuits, but it is possible to use just the inner dynamics of memristors to perform learning~\citep{Caravelli2018}.

The main question that remains unanswered is whether analog system have an advantage over digital ones at all.
A strong argument is provided by~Vergis et al., where it is shown that analog devices can be simulated with polynomial resources on a digital machine with enough resources.
We pointed out the importance of the collective properties of memristive circuits.
The dynamics of a collection of memristors interacting on a circuit can, in principle, derived from the implementation of circuit voltage and current constraints, and strongly depends on the dynamics of a single unit.
Understanding the interaction of memristors via Kirchhoff laws can in principle enable the application of memristors to a variety of computational tasks.
Below, we make this more precise using a general mapping from digital to analog computation.
We consider a differential equation of the form:
\begin{equation}
    \tder{\vec{y}}{t}(t) = f(\vec{y}(t),t), \quad y(t_0)=y_0,\quad
\end{equation}
 which describes a physical system.
In~\citep{Vergis1986}, a constructive proof based on a Euler integration method is provided, and it is shown that the amount of resources needed to simulate the system above on a digital machine is polynomial in the quantities $R$ and $\epsilon$:

\begin{equation}
    R=\max_{t_0\leq t \leq t_f} || \tder{{}^2\vec{y}}{t^2}(t)||,\; \epsilon = ||\vec{y}(t_f) - \vec{y}^*||,
\end{equation}

\noindent where $\vec{y}^*$ is the simulated system, and thus $\epsilon$ is our required precision.
Since for quantum systems $R\propto 2^N r^*$, and $r^*$ is a constant, classical computers require an exponential amount of resources to simulate a quantum physical systems.
This does not mean that classical systems can always be simulated: if the second derivative is large, our system requires a lot of computational power to be simulated on a digital machine.
A similar argument applies also to quantum computers, on which there has been a huge effort in the past decades.
In the case of a quantum system with $N$ qubits, the~vector $\vec{y}(t)$ is $2^N$-dimensional according to the Schr\"{o}dinger equation.
In a typical (analog) electronic computer,  $\tder{{}^2 \vec{y}}{t^2}$ is the derivative of the voltage.
Thus, if our circuit presents instability, it is generically hard to simulate the system.
This is specially striking in the case of chaotic behavior, which is known to emerges when a dynamical system is connected to a hard optimization problem~\citep{ErcseyRavasz2011}.
These are also arguments against the simulation of memristive system on a digital computers, which do not apply to the actual physical system performing the analog computation.

\section{Conclusions}
In the present article, we have provided an introduction and overview of memristors, both the  applications as memory, and the appealing features of memristors beyond the purpose of memory storage.
We have discussed the history of memristors, with the purpose of helping the reader understand their role in modern electronic circuitry, and why memory is a normal feature at the nanoscale.
The perspective which we have tried to provide is that, despite current applications focus on the implementation of standard machine learning algorithms on chips, memristors can be used to perform analog computation which goes beyond the standard framework of crossbars.

In magnetic materials, which are commonly used for memory purposes, the interaction between the magnetic spins (which represent the bits) is purposely screened to avoid the spins to flip, and the memory be lost. Via the interaction between the spins however, logic gates can be constructed~\citep{boospin}.
Similarly, memristors can be used as memories, thus trying to avoid their interaction in the circuit via Kirchhoff laws, or one can try to harness their interaction to perform computation.
At the most basic level, memristors can be interpreted as synapses, and the introduction of hybrid CMOS-memristor technology can allow the implementation of supervised and unsupervised machine learning algorithms that make use of these features.
This resonates with the fact that memristors have been proposed to play an important role in biology, as for instance in the case of plant signaling, amoeba adaptation, and the Hudgkin--Huxley model of squids neurons.

Building on these ideas, it make sense to pursue the complex dynamical features of memristors, interacting via Kirchhoff laws, for self-organizing computational devices.
Before the dynamical features of memristors can be fully harnessed, it is though imperative to be able to understand the single memristor physical principle in order to implement reliable memory units, in particular using the crossbar array framework we have described.
This task requires a deep knowledge of the dynamics of a single memristor component, via models which accurately describe the relevant behavior of the~device.

We have also tried to emphasize that there are several mechanisms which enable resistive switching, and different components will follow different mechanisms to change the internal state.
Despite their differences, the dynamics of memristive components share a common feature, which is the competition between two internal phenomena.
These phenomena can be cast as ``forgetting'' (decay to an off state when voltage is not applied) and ``reinforcement''  (tendency to state change which depends on the current).
As discussed by~Kohonen, this competition is an important feature among several analog computing paradigms.
As examples, we take ant-colony optimization~\citep{Dorigo1997} and experimental results with memristive devices~\citep{Ohno2011,WilliamsDD}.

We have also pointed out the importance of the collective properties of memristive circuits for computation---in particular, how we can harness the intrinsic variability of memristors to different computational problems.
However, analog machines are good at specific computational tasks, while digital ones excel for their generality.
Thus, the integration of CMOS and analog systems in future computers is favorable in the long term.
The memristor is one of many technologies which are able to encode computational tasks and simultaneously be used as memory, which can be easily integrated in modern computers.
However, in certain instances, the von Neumann architecture is not the best architecture for performing calculations, and we mentioned the case of quadratic optimization and generic combinatorial problems.

Due to our focus in memristive devices for computation, and the models of computation that are compatible with their properties, we have omitted reviewing the role that memristive system can have in biological cognitive systems, and their relation to stochastic resonance~\citep{Moss04,McDonnell09,McDonnell2011,Stotland2012,Slipko2013,Patterson2014,Fu2018,Feali2017}.
This decision was taken to keep the presentation focused and consistent.

In conclusion, we have provided an overview of the current research questions and applications of memristive technology.
We have provided also a rather long, but far from exhaustive bibliography on the subject which might help the interested reader in learning the subject.

\vspace{6pt}

\authorcontributions{F.C. and J.P.C. contributed equally to this work.}

\funding{F.C. acknowledges the support of NNSA for the U.S. Department of Energy at Los Alamos National Laboratory under Contract No. DE-AC52-06NA25396.
J.P.C. received support from the discretionary funding scheme of the Swiss Federal Institute of Aquatic Science and Technology project EmuMore.}
\acknowledgments{We thank Fabio L. Traversa, F. Sheldon, Magdalena E. Dale, Giacomo Indiveri, Alex Nugent,  Miklos Csontos, Themis Prodromakis, Yogesh Joglekar, for their useful comments and observations, and Shihe~Yang, Teuvo Kohonen, Themis Prodromakis, Yogesh Joglekar and Qiangfei Xei for the permission of using figures from their work for this review. Also, we thank in particular Magdalena E. Dale for helping us with the list of companies.}

\conflictsofinterest{The authors declare no conflict of interest.}

\appendixtitles{yes} 
\appendixsections{multiple} 

\appendix
\section{Physical Mechanisms for Resistive Change Materials}
\label{sec:appphys}
In the main text, we have mentioned that Branly's coherer can be interpreted as a granular material induced memristor. Before we discuss some more modern memristors, we believe it makes sense to give a sense why granularity is important for nonlinear resistive behavior.
Branly's coherer serves as perfect homemade memristor, as it simply requires either a (fine) metallic filling or some metallic beads, and it falls within the Physics discipline of electrical properties of granular media.
The qualitative and quantitative behavior of the case of the metallic beads contained in (and constrained to) an insulating medium of polyvinyl chloride (PVC) 
is presented below~\citep{Falcon2005,Bequin2010}.
The metallic beads are assumed to be in mutual contact, and a force $F$ applied at the two extremities enforces it.
We assume that there are two temperatures in the system: one at the microcontact between the beads at an equilibrium temperature $T$~\citep{Falcon2005} and a room temperature $T_0$, which is the one of the beads.
Without going into the details, it is possible to show a nonlinear behavior in the resistivity of $N$ beads via the study of the contact between the beads.
The metallic beads contacts can be though of as Metal--Oxide--Oxide--Metal contact, reminiscent of the memristor we will discuss below.
Kohlrausch’s equation establishes the voltage drop at the contact.
Given the current $I$ flowing in the beads, one has that
\begin{equation}
    V=N L \int_{T_0}^{T_{c}} \lambda(T) \rho_{el}(T) dT,
\end{equation}

\noindent where $\lambda(T)$ is the thermal conductivity of the material and $\rho_{el}(T)$ the density of electrons, while $T_0$ is the beads' room temperature and $T_c$ the maximum temperature at the contact when the current is flowing.
If $R_0$ is the resistance when the contact is cold, clearly we can rewrite (using Ohm's law) the equation above as the following effective equation:
\begin{equation}
    I R_0=V(T),
\end{equation}

\noindent where $R_0$ depends on the geometry of the contact.
It can be shown, however, that the maximum temperature $T_c$ depends on the voltage as $T_m^2=T_0^2+\frac{U^2}{4 L}$, where $L$ is the Lorentz constant, by noticing that via the Wiedemann--Franz that $\rho_{el} \lambda= L T$.
In addition, Mathiesen's rule for the electron mobility shows that the electron mobility is linear in $T$, with a proportionality constant that is material dependent.
Putting all these facts together, it is not hard to see the hysteretic behavior of the system.
This~effective model reproduces well the controlled experiments.
Since the voltage drop is zero when the current is zero, this also implies a pinched hysteresis, or resistive behavior.
These ``mechanical'' nonlinear resistors are prototypes for the more complicated case shown below~(see \citep{Falcon2005} for more details).
A~posteriori, many phenomena, both quantum and classical, have been reinterpreted as ``memristors''  (for instance, Josephson junctions~\citep{supercm}).
Below, we provide a non-exhaustive list of mechanisms that have been discussed in the literature. For more detailed information, we suggest the reviews on the subject of resistive switching given in~\citep{DiVentra2011,Kuzum2013,Yang2013,Mikhaylov2016}.

\subsection{Phase Change Materials}
\begin{figure}[H]
    \centering
    \includegraphics[scale=2]{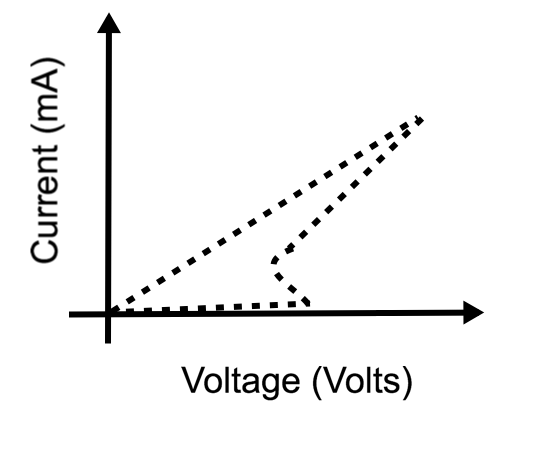}
    \caption{Current--Voltage diagram typical of a phase change materials (PCM) memory storage. The sharp transition at low currents and at a typical ``Critical voltage'' is clearly visible.}
    \label{fig:PCM}
\end{figure}

Phase change materials (PCM) are glassy materials: this implies that usually these materials can have different phases in which the material can be either ordered (crystalline-like) or disordered (amorphous, as in a liquid).
In these two phases, we associate two different resistances.
If the structural change can be associated to the values of an applied voltage, then when the transition occurs one has a rather quick transition from one resistive state to another~\citep{Ovshinsky1968,Neale70,Buckley1975} due to an electrical instability.
These~materials are considered memristors by some, but not by all researchers, and were discovered as early as the late 1960s~\citep{Ielmini2011} in amorphous chalcogenides.
This is the most mature of the emerging memory technologies.
Since we have used various analogies before, the reader in need of a visual way to understand these type of materials might find some ideas on the shelves of a pharmacy.
Phase-change materials are being used for instant freeze packages normally used in case of injuries, and are also called ``gel packs''.
In order to initiate the cooling, users typically need either to mix two materials, or quickly apply a certain force on the package.
The material will use the ``kick'' to initiate a phase transition, absorb the heat and thus lower the temperature of the package.
PCM memory devices work in a similar manner, but on a much lower scale ($\sim$20 nm), and the ``kick'' is provided by the electric field. These materials were not commercial for years due to the rapid advancement of silicon-based technology.
The typical I--V diagram is shown in Figure~\ref{fig:PCM}.
One generically has two types of materials squeezed between two electrodes: on one side, one has an insulating material with a small conducting channel, directly connected to the phase change material.
As the channel heats up, the~phase change material locally changes phase starting from point of contact at the conduction channel, until it reaches the other electrode.

In some chalcogenides-based memristors (called Self-Directed channel memristors, or SDC), ions are constrained to follow certain channels, and their operation is similar in many ways to both PCM-components and atomic switches~\citep{campbell}.

\subsection{Oxide Based Materials}
\begin{figure}[H]
    \centering
    \includegraphics[scale=2]{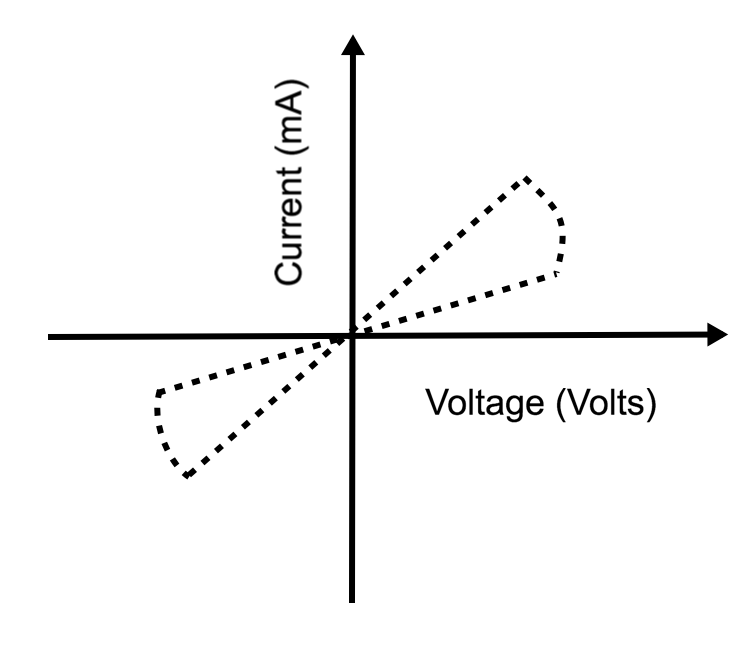}
    \caption{Bipolar and non-volatile switching pinched hysteresis.}
    \label{fig:bipnonv}
\end{figure}

The oxide and anionic materials based on transition metal work instead differently.
In order to understand why a memristor might be different from a normal conductor, it is useful to understand what happens when two materials with different properties are ``merged'' together: those of charge donor (excess of electrons) and charge receiver (excess of electron ``holes''), which are also called doped and undoped materials.
In oxide materials, the carrier of the charge is typically the oxygen.
It should be mentioned that, whilst various mechanisms have been suggested, very likely all of these coexist in a typical oxide material, including filament formation~\citep{Hoskins2017}.
In general, whether the resistive switching is thermally or electrically driven, the typical understanding is that a chemical transition occurs in the material, and that the hysteresis is due to vacancy movements in the materials.
The transition is either directly driven by the direct application of the electric field, or as a byproduct of the heating of the material due to the current.
In semiconductors, the key quantity of interest is the energy gap $E_g$ between the valence and the conduction channels.
If the gap is of the order of $E_p \approx k T$, then thermal effects which might let a charge carrier jump into the conduction channel become important.
If the gap is too large, electric effects are the dominant ones.
One example is provided by the bipolar and non-volatile switching, described by  Figure~\ref{fig:bipnonv}.

\begin{figure}[H]
    \centering
    \includegraphics[scale=1.5]{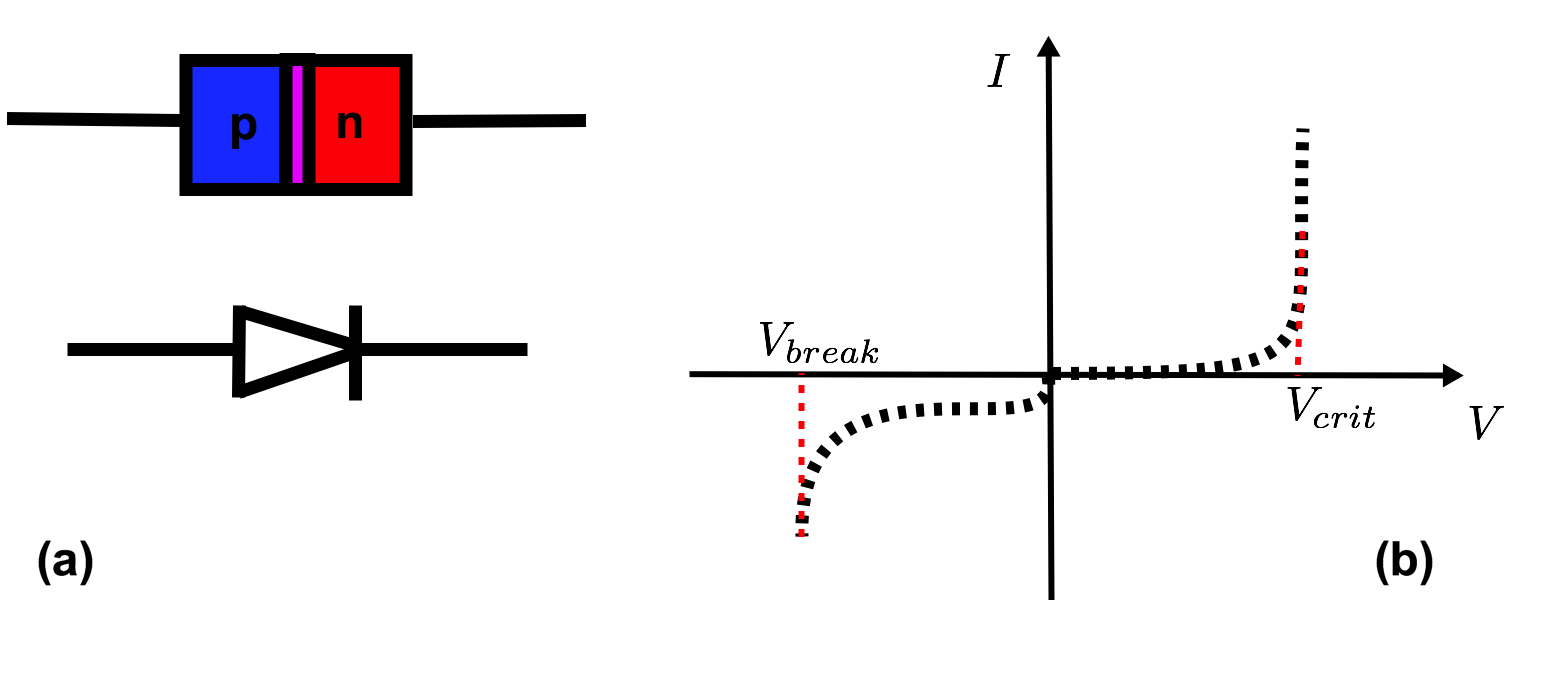}
    \caption{The pn-junction. (\textbf{a}) the pink region is the charge exchange region for the doped and undoped region; (\textbf{b}) a stylized response in voltage of a diode. In many ways, the shape resembles the nonlinearity which occur in nonlinear memristors in which field effects become dominant.}
    \label{fig:pnj}
\end{figure}

The shape of I--V curve generically shows what type of mechanism is underlying the switching.
As~the field effects become more prominent (for instance, the effect of Schottky barriers at the junctions), the I--V diagram becomes nonlinear.
There can be other non-volatile and non-volatile switching in which we are not discussing here, and due to thermal excitation only.
A simplified model of vacancy-charge movement in the dielectric has been proposed in~\citep{Chernov2016}.
When nonlinearities are present in non-volatile materials, it means that the switching is dominated
by the electrical switching.
In many ways, some of the physical phenomena happening in memristors can be intuitively understood in terms of the simplest semiconductor: the diode, represented in Figure~\ref{fig:pnj}a.

The diode can be thought of as a dramatically nonlinear resistance, made by merging two materials, a doped (filled with defects) and undoped one, with a thin interstitial when charges exchange.
We~can characterize the diode by the two voltages $V_{break}<0<V_{crit}$.
For voltages above $V_{crit}$, the diode will have a very low resistance, and, for voltages lower  than $V_{break}$, the diode will breakdown.
In many ways, the shape of Figure~\ref{fig:pnj}b is what is usually seen in memristive oxide components, with the difference that the memristor will exhibit a hysteresis.
At the interfaces (pink zone in Figure~\ref{fig:pnj}a), the~charge carriers will have to overcome a barrier (Schottky barrier) (characterized by $V_{crit}$) to continue their flow into the material.
When this barrier is overcome, the flow is almost free.
On the other hand, when the flow is inverted, the undoped material will act as a very high barrier to the flow into the dielectric region.
However, the charge carriers can still penetrate the material and damage it in an irreversible manner.
These are usually called  Zener cascades, and are often observed in memristors.
Albeit cartoonish, this picture can help understand some of the nonlinear phenomena happening in memristors, and not captured by the linear resistance model we discussed above.
As a final comment, endurance in oxide materials is one of the key problems in the technological competitiveness of memristors~\citep{IelminiEndurance2015}.
When the number of cycles becomes comparable with the lifetime of the component, some exotic phenomena as multiple pinchpoints (tri- or four-lobes hysteresis loops) in the V--I diagram can also occur.
These can be modeled via the introduction of fractal derivatives~\citep{8351761}.

\subsection{Atomic Switches}
\begin{figure}[H]
\begin{subfigure}{0.5\textwidth}
    \centering
    \includegraphics[width=0.6\textwidth]{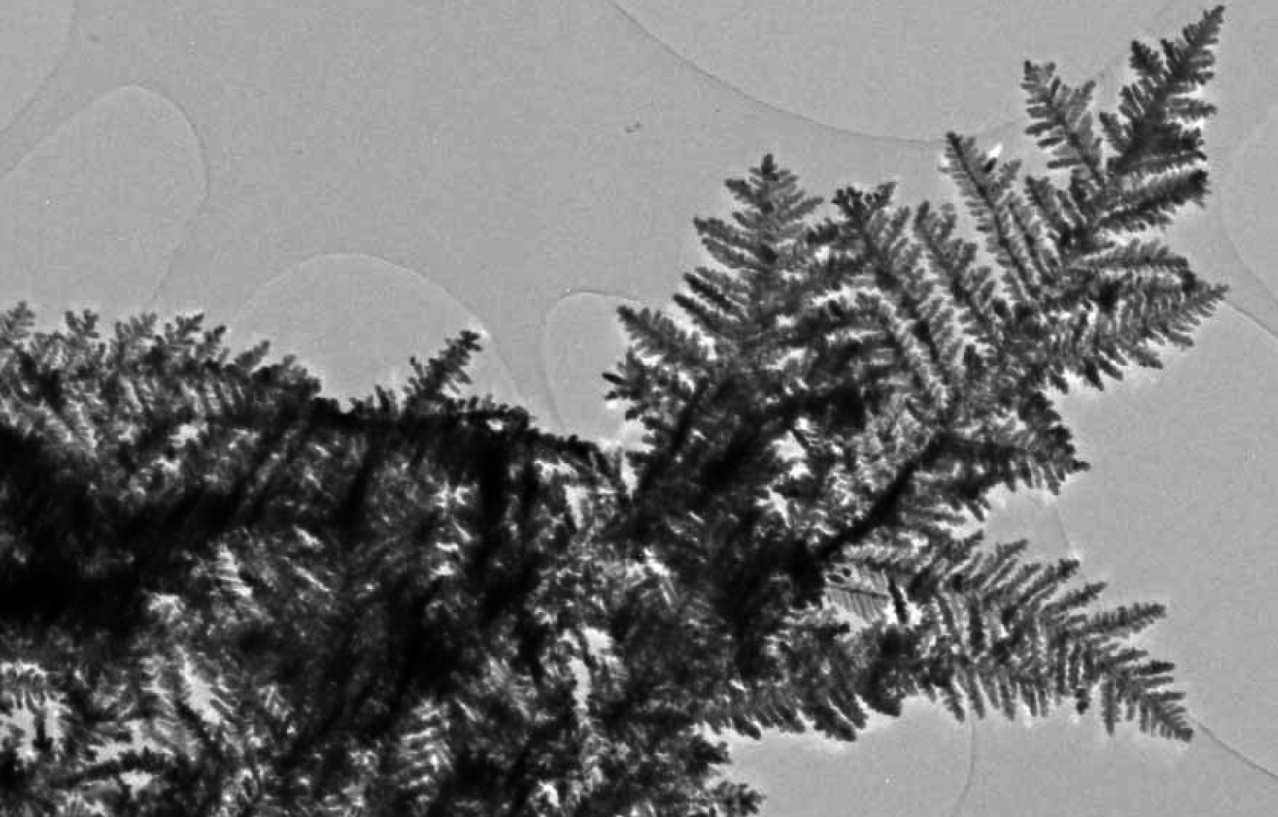}
    \caption[]{} 
    \label{fig:dendr}
\end{subfigure}
\begin{subfigure}{0.5\textwidth}
    \centering
    \includegraphics[width=0.5\textwidth]{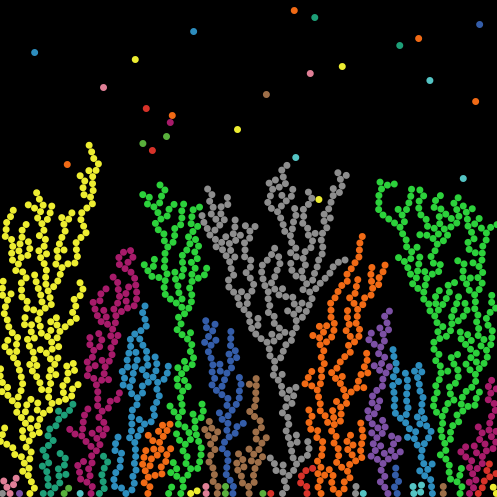}
    \caption[]{}
    \label{fig:dla}
\end{subfigure}
\caption{Comparison between dendritic growth in silver ion materials and Diffusion-Limited- Aggregation simulated using NetLogo. (\textbf{a}) dendritic growth in silver ion at the micro meter scale, reproduced with permission from~\citep{Wen2006}; (\textbf{b}) diffusion-limited-aggregation simulated using NetLogo.}
\label{figA4}
\end{figure}

The third mechanism described in this paper (albeit not the last!), and maybe the most visually appealing, is the filament growth memristor. 
The materials, often called atomic switch networks in the literature, work in a slightly different manner than the previously described memristors~\citep{Stieg2014,Sillin2013,Avizienis,Stieg12,WilliamsDD}.
The idea behind these components is that the two electrodes work, when the electrical field is applied, as an electrode and a cathode.
The applied electric field induces an electrochemical reaction which triggers the growth of filaments.
From a highly resistive state, these filaments reduce the resistivity by introducing new channels for the charge carriers to flow.
The closest physical growth model which can describe the growth of the filaments is provided by Diffusion-Limited-Aggregation (as in Figure~\ref{figA4}b).
Each colored filament represents one possible channel. The typical charge carriers are either silver ions or some silver compounds.
There are not many experiments focused instead on the collective behavior of memristive networks. It is worth mentioning, however, what are the observed features for $Ag^+$, or Atomic Switch Networks, whose collective dynamics is interesting for the emergence of seemingly critical states~\citep{Scharnhorst2018}.
In fact, whilst the dynamic of a single filament is simpler to describe, the system exhibits collective power law power spectrum with an exponent close to 2.
Albeit this exponent can be explained via the superposition of wide range of relaxation timescales for each memristor~\citep{Caravelli2017}, the~critical behavior is the accepted one because of their intrinsic nonlinearity~\citep{Avizienis}.
In order to see the variety of memristive behavior, we consider two models here suggested in the literature which are different than the simple TiO$_2$ linear model memristor.

The first, suggested in ~\citep{Pickett2009} as a phenomenological switching model between two off and on states in TiO$_2$, is of the form:
\begin{align}
R(w) &= \Roff (1-w)+\Ron w,\\
\tder{w}{t}&= \begin{cases}
f_{\text{off}} \sinh(\frac{i}{i_{\text{off}}}) e^{-e^{\frac{w-a_{\text{off}}}{w_c}-\frac{|i|}{b}}-\frac{w}{w_c}} & i>0, \\
f_{\text{on}} \sinh(\frac{i}{i_{\text{on}}}) e^{-e^{-\frac{w-a_{\text{on}}}{w_c}-\frac{|i|}{b}}-\frac{w}{w_c}} & i<0,
\end{cases}
\end{align}
\noindent where the parameters $f_\text{on/off}$, $i_\text{on/off}$ and $a_\text{on/off}$ are state dependent, while $w_c$ and $b$ are not.
In addition, the~model above shows that the energy depends exponentially on the current.

Another metal oxide of interest is $WO_x$, studied in~\citep{Chang2011}.
The model equations for a single component are given by:
\begin{align}
    I&=\alpha (1-w) \left(1-e^{\beta V} \right) + w\gamma \sinh(\delta V), \\
    \tder{w}{t}&= \lambda \left(e^{\eta_1 V}-e^{-\eta_2 V} \right).
\end{align}

The model above, it is interesting to note, is controlled in voltage rather than current, and the parameters $\alpha,\beta,\gamma,\delta$ and $\eta_{1}$ and $\eta_{2}$ are positive.
For $V=0$, $I=0$, which implies a pinched hysteresis (however, the hysteresis is different from TiO$_2$ devices).
The parameter $w$ is physically interpreted as the portion of the device in which oxygen charges tunnel through the device.
For $w=1$, one has a tunneling dominated device, while, at $w=0$, one has a Schottky-dominated conduction.

One question which might be relevant to mention at this point is: what is the advantage of using memristors rather than other memory devices?
The perspectives of the present paper is that there are two different reasons why these devices can turn useful~\citep{itrs11}.
The first advantages is the density compared to standard memory.
For instance, compared to DRAM and static random-access memory (SRAM), memristors (or PCM) retain memory for years rather than less than seconds. 
Compared to SRAM, however, whose read-write time is less than a nanosecond, memristors with current technology are one order of magnitude slower.
In addition, in terms of read--write cycles, the technology of memristors and PCM is between 3--5 orders of magnitude less, but still much more durable than hard disk drive (HDD). 
The picture is that memristors and PCM are not uniquely better than the current standard in computing.
\subsection{Spin Torque}
Spin-torque memory materials ~\cite{stt} have an advantage in terms of durability~Grollier et al. over other materials such as transition metal oxides.
These are often considered as second-order memristive devices but a simplified model of spin-torque induced resistance is provided in~\citep{XiaobinWang2009}.
The starting point is the Landau--Ginzburg--Gilbert (LGG) equation with a spin-torque interaction~\citep{Sun2000}. We consider two magnetic layers perpendicular to the flow of the current, and in which one is fully polarized: its magnetic orientation is fixed in a direction perpendicular to the current, while the second layer is free. Via the LGG equation with rotational symmetry, the dynamics of the angle between the pinned and free layer is given by
\begin{equation}
    \tder{\theta(t)}{t}=\alpha \gamma H_k \sin\left(\theta(t)\right) \left(p-\cos \left(\theta(t)\right)\right),
\end{equation}
where $\gamma$ is called gyromagnetic ratio, $\alpha$ is the damping parameter, $H_k$ is the perpendicular anysotropy in the free layer and $p=f \hbar I$ is a current dependent which represents the effect of the current-polarization interaction. We have emphasized the presence of $\hbar$ to imply that this is a purely quantum correction.

In order to see how this device is a memristor, we see that, in the simpler case with full rotational symmetry, one has an induced magneto-resistance $R(\theta)$, which depends on the angle $\theta$ in the case of full rotational symmetry as:
\begin{equation}
    R\left(\theta(t)\right)=\frac{1}{a +b \cos\left(\theta(t)\right)},
\end{equation}
where $a$ and $b$ are constants which depend on the resistance in the free layer and on the ration between the highest and lowest achievable resistances. We thus see that, in the case of full rotational symmetry, spin-torque materials are first order memristors.

\subsection{Mott Memristors}
In this section, we discuss resistive switching due to the Mott between insulating and conducting phase in metals~\citep{mottmemr,HopMemr2}.
In the Mott transition, the important quantity of interest is the electron density, which acts as a control parameter for the transition.
This implies that, in general, the transition can be induced via pressure change or doping.
The transition is usually studied in the context of weakly correlated electron liquids, in which the Coulomb interactions are screened and thus can be considered weak.
In general, there are various known mechanisms that enable an insulator transition.
At the single electron (charge carriers in general), there can be band insulators, which we have discussed before in the context of oxide based materials---Peierls insulators~\citep{peierlst}, which are due to lattice deformations that distort the band gaps; or Anderson insulators~\citep{andersont}, an effect due to the interaction of a particle with the disorder in the materials in which, at the quantum level, the charge carrier becomes localized.

The Mott transition occurs due an electron--electron interaction, and leads to the formation of a gap between the ground state of the system and the excited state: this implies that, in order to kick a charge into the conduction band, the system requires overcoming a barrier.
The transition occurs because of the repulsion between the electrons, which impedes the free flow of these into the material.
Mott memristors take advantage of the fact that certain materials (such as niobium dioxide~\citep{niobium}) have a current induced Mott transition.
This is due to the fact that, as the current is enough to locally heat the material above a certain threshold value, the material undergoes phase transition.

\section{Sparse Coding Example}
\label{sec:scod}

Sparse coding is the solution of the following optimization problem:
\begin{align}
\vec x&=\sum_{i=1}^M a_i \vec \phi^i, \\
\text{min}_{\vec a}& \Vert\vec a\Vert_0,\ \Vert\cdot\Vert\text{ 0-norm},
\end{align}

\noindent and which is notoriously NP-hard.
Given some technical conditions which we do not discuss, the problem above can be approximated in certain situations by replacing the 0-norm with the 1-norm, and the minimization replaced with
\begin{equation}
    \text{min}_{\vec a}\Vert\vec x-\sum_{i=1}^M a_i \vec \phi^i\Vert_2^2+ \lambda  \Vert\vec a\Vert_1,
\end{equation}

\noindent where $\Vert\cdot\Vert_2$ is the two norm and $\lambda$ a Lagrange multiplier.
The problem above can be encoded in a neural system via a locally competitive algorithm (LCA), which is formulated as follows.
Given the coefficients $a_i(t)$, consider a ``neuron'' variable $u_i(t)$ such that $a_i(t)=T_{\lambda}\left(a_i(t)\right)$ and where $T_\lambda(\cdot)$ is a threshold function.
We consider an energy function of the form
\begin{equation}
    E=\frac{1}{2} \Vert\vec x -\hat x\Vert+ \lambda \sum_m C(a_m),
\end{equation}

\noindent with $\vec x$ defined as above and $\hat x=\sum_{i=1}^M a_i \vec \phi^i$, and $C(\cdot)$ a certain unspecified cost function.
One then looks at a dynamics for the neuron state $u_i(t)$ of the form
\begin{equation}
   \dot u_i(t)= \frac{1}{\tau} \frac{\delta}{\delta a_i} E,
\end{equation}

\noindent for a certain relaxation constant $\tau$, which it can be easily seen to be defined, given $\vec b(t)=\Phi \vec x(t)$, $\Phi=[\vec \phi^1,\cdots,\vec \phi^M]$, as 
\begin{equation}
    \dot u_m(t)=\frac{1}{\tau}\left(b_m(t)-u_m(t)-\sum_{n\neq m} G_{mn} a_n(t) \right),
\end{equation}

\noindent with $G_{mn}=\vec \phi^m\cdot \vec \phi^n$.
The correspondence between the threshold function and the cost function is given by the equation:
\begin{equation}
    \lambda \frac{d}{da_m} C(a_m)=u_m-a_m=\left(u_m-T_\lambda (u_m)\right).
\end{equation}

The equations above are suitable to be implemented on a memristive circuit, as besides a forcing term and a leaky integration term, there is a nonlinear integration term and are implementable via Hopfield continuous networks~\citep{Hopfield1,Hopfield2}.
The equations above would be linear if the thresholding function $T_\lambda$ were trivial.




\reftitle{References}

\end{document}